# Paying for Privacy: Pay-or-Tracking Walls

**Timo Müller-Tribbensee[1], Klaus M. Miller[2], Bernd Skiera[3]**

**March 5, 2024**

COMMENTS WELCOME

[1] Timo Müller-Tribbensee, Doctoral Candidate, Department of Marketing, Faculty of Economics and Business, Goethe University Frankfurt, Theodor-W.-Adorno-Platz 4, 60323 Frankfurt, Germany, Phone +49-69-798-34565, mueller-tribbensee@wiwi.uni-frankfurt.de.

[2] Klaus M. Miller, Assistant Professor of Marketing, HEC Paris, Rue de la Libération 1, 78350 Jouy-en-Josas, France, Hi!PARIS Chairholder, Center in Data Analytics & AI for Science, Technology, Business & Society, millerk@hec.fr.

[3] Bernd Skiera, Professor of Electronic Commerce, Department of Marketing, Faculty of Economics and Business, Goethe University Frankfurt, Theodor-W.-Adorno-Platz 4, 60323 Frankfurt, Germany, Phone +49-69-798-34649, skiera@wiwi.uni-frankfurt.de; also Professorial Research Fellow at Deakin Business School, 221 Burwood Highway, Burwood, VIC 3125, Australia, bernd.skiera@deakin.edu.au.

Financial Disclosure:

This project has received funding from the European Research Council (ERC) under the European Union's Horizon 2020 research and innovation program (grant agreement No. 833714).

# Paying for Privacy: Pay-or-Tracking Walls

**Abstract**

Prestigious news publishers, and more recently, Meta, have begun to request that users pay for privacy. Specifically, users receive a notification banner, referred to as a pay-or-tracking wall, that requires them to (i) pay money to avoid being tracked or (ii) consent to being tracked. These walls have invited concerns that privacy might become a luxury. However, little is known about pay-or-tracking walls, which prevents a meaningful discussion about their appropriateness. This paper conducts several empirical studies and finds that top EU publishers use pay-or-tracking walls. Their implementations involve various approaches, including bundling the pay option with advertising-free access or additional content. The price for not being tracked exceeds the advertising revenue that publishers generate from a user who consents to being tracked. Notably, publishers' traffic does not decline when implementing a pay-or-tracking wall and most users consent to being tracked; only a few users pay. In short, pay-or-tracking walls seem to provide the means for expanding the practice of tracking. Publishers profit from pay-or-tracking walls and may observe a revenue increase of 16.4% due to tracking more users than under a cookie consent banner.

*Keywords*: privacy, tracking, consent, behavioral targeting, online advertising

In May 2018, the Austrian newspaper "Der Standard" became the first prominent European publisher to ask its users to pay to avoid being tracked online. Its innovation centered around a notification banner for first-time users of a website or an app that offers two options: (i) pay money to avoid being tracked or (ii) consent to being tracked, which involves collecting and processing the user's browsing behavior for, e.g., behaviorally targeted advertising. If users choose neither option, they will be unable to access the content. We refer to this practice as a pay-or-tracking wall, which practitioners also refer to as a "cookie paywall", "cookie wall with paid alternative", "accept-or-pay cookie banner", or "pay-or-okay" banner.

Other European publishers have followed suit, including prestigious news websites such as "Der Spiegel" in Germany (see Figure 1a)) and "Le Monde" in France. Even Meta recently implemented a pay-or-tracking wall for European users of Facebook (see Figure 1b)) and Instagram (Meta 2023). The spread of this approach has led some activists and users to worry that Internet privacy might soon become a luxury of the affluent, accessible only through payment. Conversely, publishers are hoping to produce revenue in the wake of the European General Data Protection Regulation (GDPR), which typically requires user permission for tracking and has made behaviorally targeted advertising more challenging.

Evaluating those concerns is difficult because we know little about pay-or-tracking walls and publishers' usage of them. Morel et al. (2022) first reported on the prevalence of pay-or-tracking walls, while subsequent work by Morel et al. (2023) and Rasaii et al. (2023) detected an increasing adoption. However, scholars remain unclear on the extent to which top publishers have opted for this approach and how varied the implementations are. In addition, we do not fully grasp users' reactions to this practice or the benefits that publishers derive.

*Figure 1: Exemplary Pay-or-Tracking Walls.*

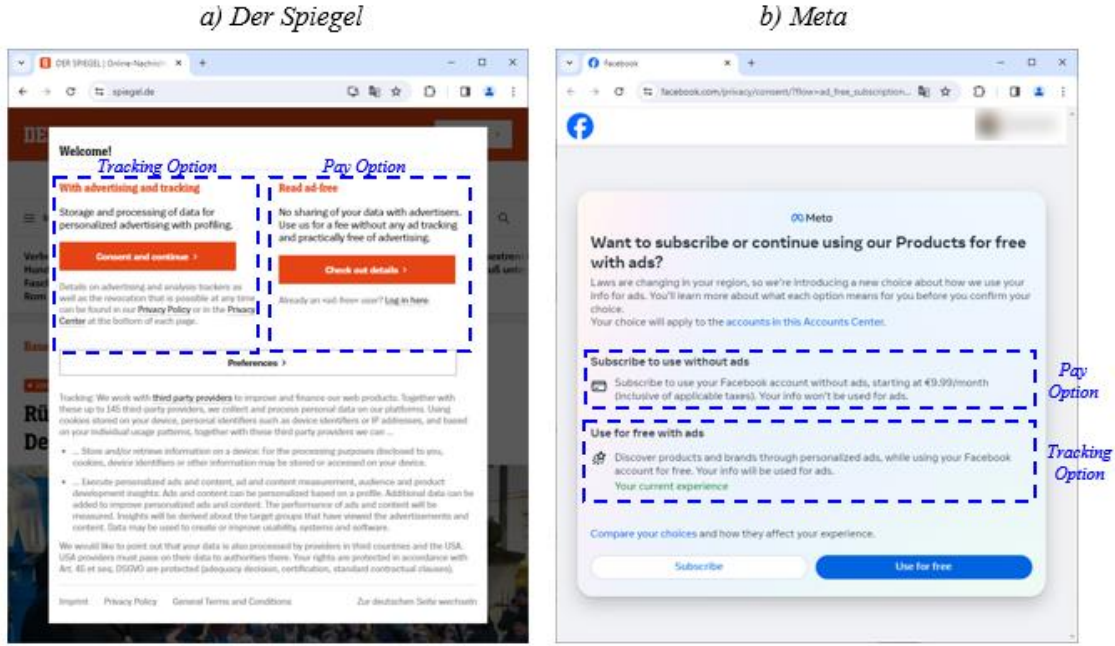

*Notes*: Snapshots from January 8, 2024, taken on the websites' English versions. Blue dashed lines and labels added by authors. "Der Spiegel" and "Meta" bundle pay for not being tracked with advertising-free access.

Publishers stand to potentially profit from substantial gains through the "pay option" or from many users choosing the "tracking option". While the former indicates that users are willing to pay for their privacy, the latter enables publishers to sell ad inventory at higher prices thanks to behaviorally targeted advertising (e.g., Marotta, Abhishek, and Acquisti 2019). That said, publishers may face losses if many users decide on the "leave option" in response to a pay-or-tracking wall.

Against that background, this article aims to explore and understand publishers' use of pay-or-tracking walls, users' reactions, and the resulting economic consequences. More specifically, we answer the following research questions (RQ):

- Which publishers use pay-or-tracking walls? (RQ1)

- How do publishers implement pay-or-tracking walls? (RQ2)

- How do users react to pay-or-tracking walls? (RQ3)

- What are the economic consequences of pay-or-tracking walls? (RQ4)

We conducted four empirical studies using four different datasets. While the first two studies address RQ1 and RQ2 by exploring the popularity and implementation of pay-or-tracking walls, the third and fourth studies investigate users' reactions to address RQ3. Lastly, we combine the insights from the four studies to outline the economic consequences of implementing a pay-or-tracking wall, addressing RQ4.

Our findings reveal that top publishers in Austria, France, Germany, and Italy use pay-or-tracking walls in diverse implementations, such as bundling the pay option with advertising-free access or additional content. We observe that the price for not being tracked exceeds the advertising revenue that publishers generate from users who consent to being tracked, indicating that users seeking privacy have to pay a premium. Regarding users' reactions, publishers' traffic does not decline when implementing a pay-or-tracking wall; most users consent to being tracked and only a few users pay. Thus, compared to its alternative (i.e., a cookie consent banner allowing users to refuse being tracked at no cost), publishers profit from pay-or-tracking walls and may observe a revenue increase of 16.4% due to tracking more users than under a cookie consent banner.

Our results have implications for multiple stakeholders, including publishers planning to implement a pay-or-tracking wall, users seeking to better understand the approach, and regulators assessing the associated advantages and risks. Since publishers profit from introducing pay-or-tracking walls, our findings suggest that publishers should adopt them. However, the results also lend weight to concerns that online privacy may become a luxury for the affluent, as privacy-sensitive users will have to focus on a few publishers if they do not want to spend too much money. From a regulatory perspective, the low share of users choosing to pay raises questions about compliance and whether the European privacy framework has actually improved users' data control. Instead, pay-or-tracking walls seem to provide the means to expand the practice of tracking.

The paper contributes to the literature on monetizing digital content (e.g., Aral and Dhillon 2021; Cao, Chintagunta, and Li 2023; Pattabhiramaiah, Sriram, and Manchanda 2019). It also adds to several domains in the privacy literature, such as users' privacy preferences (e.g., Acquisti, John, and Loewenstein 2013; Tsai et al. 2011), the privacy paradox (e.g., Athey, Catalini, and Tucker 2017), privacy-enhancing technologies (e.g., Brough et al. 2022; Yan, Miller, and Skiera 2022), and the economic effects of privacy regulations (e.g., Johnson, Shriver, and Du 2020; Peukert et al. 2022). In particular, this study contributes to the recent research agenda outlined by Tucker (2023), including examining users' taste for privacy, the market for privacy-enhancing technologies, and the connection between privacy and inequality.

## Description of Pay-or-Tracking Walls

### *Definition of a Pay-or-Tracking Wall*

A pay-or-tracking wall is a notification banner displayed on the publisher's property, such as its website or app. It requests that users either pay money to avoid being tracked ("pay option") or consent to being tracked ("tracking option"), the latter of which involves collecting and processing users' browsing behavior over time. If they choose neither option, users will be unable to access the content and will have to leave.[1]

Granted, this is not the only approach that firms use to obtain tracking permission. Compared to the three options offered by a pay-or-tracking wall (i.e., pay, consent to being tracked, or leave), the tracking wall (Zuiderveen Borgesius et al. 2017) only provides two options (i.e., consent to being tracked or leave). More lenient approaches, such as Apple's ATT prompt (Kesler 2023) or typical cookie banners (Degeling et al. 2019), provide a different set of three options: namely, refuse to be tracked (at no cost), consent to being tracked, or leave.

---

[1] An exception is the French website lequipe.fr, which does not require the user to leave. Instead, the notification banner overlays approximately two-third of the screen, which makes accessing the content practically impossible.

The main novelty of pay-or-tracking walls, "pay to not be tracked", can be bundled with other features. For example, publishers could offer fewer ads or more (free) content with the pay option. Furthermore, the publisher could also just offer less tracking instead of no tracking.

### Design of Pay-or-Tracking Walls

To gain insight into how publishers implement pay-or-tracking walls, we first outline publishers' design choices, which will provide a framework for the later empirical analysis. There are four design elements that comprise the major variations between pay-or-tracking walls: tracking, advertising, content, and price. Most importantly, publishers must define the differences between the pay and tracking options (see Table 1). We elaborate on each dimension below.

*Table 1: Design Choices for Pay-or-Tracking Walls*

| Dimension | Difference Between Pay and Tracking Options | Design Choice |
|---|---|---|
| Tracking | Pay < Tracking<br>*("less tracking in pay")* | – Tracker usage and reduction |
| Advertising | Pay ≤ Tracking<br>*("equal or less advertising in pay")* | – Bundling with advertising-free access |
| Content | Pay ≥ Tracking<br>*("equal or more content in pay")* | – Bundling with additional content |
| Price | Pay > Tracking<br>*("higher price for pay")* | – Price level and difference<br>– Price model (e.g., pay-per-use / pay-per-period)<br>– Payment-related data disclosure |

*Design Choices Regarding Tracking*

Publishers need to examine and reflect on their implementation of so-called trackers, which refer to a piece of software (e.g., a cookie or pixel) that is embedded in the website or app (see Bujlow et al. (2017) for a list of tracking technologies). A tracker can collect user information and browsing behavior, such as device characteristics, visited websites, or clicks. Besides using their own trackers (first-party tracking), publishers rely on trackers from third-party providers (third-party tracking). The latter transfer information to the tracker providers, enabling publishers to integrate additional functionalities, including advertising, analytics, social network integration, and personalization (Mayer and Mitchell 2012). For instance, publishers might incorporate the DoubleClick tracker (from Google) to manage ad sales or the Chartbeat tracker (from Chartbeat) to gain audience insights.

Publishers select the set of trackers they use in the tracking option and the degree to which their pay option reduces the use of trackers. For instance, publishers could use a variety of trackers in the tracking option while entirely omitting tracking in the pay option. Conversely, publishers could selectively curtail specific tracking activities for the pay option.

*Design Choices Regarding Advertising*

Publishers typically monetize the tracking option by providing advertisers with information that enables behaviorally targeted advertising (Ada, Abou Nabout, and Feit 2022). Advertisers will typically pay higher prices to display targeted instead of non-targeted ads (e.g., Johnson, Shriver, and Du 2020; Laub, Miller, and Skiera 2023; Wang, Jiang, and Yang 2024). Conversely, publishers may exclude tracking for advertising purposes in the pay option, which means advertisers cannot conduct behaviorally targeted advertising with paying users. However, advertisers could still use other forms of advertising that do not rely on (third-party) tracking, such as contextual targeting (i.e., only information about the context or content in which the ad will occur). Alternatively, publishers could incentivize users to choose the pay option by reducing advertising: for instance, by bundling the pay option with advertising-free access to the content.

*Design Choices Regarding Content*

Design variation can also result from bundling the pay-or-tracking wall with additional (paid) content. Publishers often implement paid content such that users can choose between a free plan with limited access to content and a paid plan with unlimited access (e.g., Pattabhiramaiah, Sriram, and Manchanda 2019; Pauwels and Weiss 2008). Without paid content, the design of a pay-or-tracking wall is straightforward: Both the pay and tracking options will allow unlimited access to content. Similarly, a publisher may use paid content offerings, but not bundle them with the pay-or-tracking wall. Instead, they will treat the access plans for paid content and the pay-or-tracking wall as independent offerings with separate prices.

Alternatively, the publisher can link the pay-or-tracking wall to the paid content through two bundling strategies: price and product bundling. While the former refers to selling two separate products at a discount ("price bundling"), the latter describes selling two products as an integrated package ("product bundling"; Stremersch and Tellis 2002). We illustrate both in the context of a pay-or-tracking wall with the following example (see Figure 2).

Suppose a publisher implements two access plans: (a) a limited access plan for paid content (e.g., a metered paywall with a restriction, such as 20 articles per month, or a premium paywall with a restriction to a certain type of content, such as news articles) and (b) an unlimited access plan for paid content. Combining the two access plans with a pay-or-tracking wall (1: tracking option; 2: pay option) leads to four combinations: (a1) tracking option + limited access, (a2) pay option (for no tracking) + limited access, (b1) tracking option + unlimited access, and (b2) pay option + unlimited access. The publisher could, for instance, offer a discount if users choose combination (b2) (i.e., price bundling occurs), which is cheaper than purchasing the two (components a2 and b1) separately.

The second possibility is that the publisher uses product bundling by eliminating some combinations. For instance, the publisher could sell the pay option and the unlimited access plan only as a bundle (i.e., as the combination b2), but not separately (i.e., not as combinations a2 or b1). Alternatively, the publisher could offer either the bundle (i.e., combination b2) or the pay option with limited access (i.e., combination a2), but not the tracking option with unlimited access (i.e., combination b1). In other words, the tracking option would only be available in combination with the limited access plan.

*Figure 2: Bundling in the Context of Pay-or-Tracking Walls and Paid Content Offerings*

*Notes*: The example illustrates a paid content offering with a limited and unlimited access plan.

*Design Choices Regarding Price*

Even if publishers typically offer the tracking option at a price of 0, they must set the price of the pay option. One possibility is that publishers set the pay option's price so that it compensates for the revenue that would otherwise be generated by the tracking option. Another possibility would entail exploiting users' willingness to pay in order to maximize profit.

The price would also depend on whether publishers adopt a pay-per-use or pay-per-period plan. The former refers to charging a price per unit (e.g., per visited webpage), while the latter involves a fixed fee within a pre-defined contract period (e.g., monthly flat-rate; Rußell et al. 2020). Additionally, publishers must implement a payment process and access control for the pay option (e.g., log in via account based on email address) and may differ in the degree to which they protect users' data disclosure caused by the payment and authentication.

### Compliance of Pay-or-Tracking Walls

When requesting user permission for tracking, publishers must comply with the consent rules of the applicable privacy law. In Europe, Article 4(11) of the GDPR defines consent as *"[…] any freely given, specific, informed and unambiguous indication of the data subject's wishes by which he or she, by a statement or by a clear affirmative action, signifies agreement to the processing of personal data relating to him or her"*. Only in exceptional cases may publishers use trackers without user permission, such as for strictly technical reasons (e.g., Austrian Data Protection Authority 2023; French Data Protection Authority 2020; German Data Protection Authority 2022).

The major difference between pay-or-tracking walls and other approaches to obtaining tracking permission involves the condition of *freely given* consent. This condition requires (European) users to have a free and genuine choice and not to feel compelled to consent (Recital 42, GDPR). Legal authorities, such as the European Data Protection Board (2020), believe publishers cannot enable a free choice if they give users only two options (i.e., to either consent to being tracked or leave), which renders so-called tracking walls an illegal practice. Other approaches, such as Apple's ATT prompt or typical cookie consent banners, ensure a free choice by additionally providing the third alternative to refuse tracking, which comes at no cost. Pay-or-tracking walls offer the pay option to avoid tracking as the third alternative, but users must pay. Since a payment can put users under financial pressure, privacy activists criticize the compliance of pay-or-tracking walls with *freely given* consent (NOYB 2021).

The legal status of pay-or-tracking walls is not entirely clear; there have been no decisions by the EU's European Court of Justice to clarify case law. Some national data protection authorities (e.g., the Austrian, Danish, French, and German) have already signaled their acceptance of pay-or-tracking walls. In their published statements about pay-or-tracking walls, they have emphasized that the price of the pay option must be reasonable (i.e., not too high) to ensure a free choice (e.g., Austrian Data Protection Authority 2023; Danish Data Protection Authority 2023; French Data Protection Authority 2022; German Data Protection Authority 2023). However, regulators have not precisely established the meaning of a reasonable price. Privacy activists believe that the price should not be tethered to profit maximization, but rather the advertising revenue lost due to no tracking (NOYB 2021). Thus, when implementing a pay-or-tracking wall, publishers must consider both the regulatory environment and, remarkably, the price of their pay option.

**Aim and Overview of Empirical Studies**

As illustrated above, publishers have several options when designing a pay-or-tracking wall, and users can react differently. The following four empirical studies, summarized in Table 2, are intended to clarify publishers' and users' actual behaviors. The first study examines RQ1 by analyzing publishers' implementations, covering the popularity of pay-or-tracking walls among top European publishers and the type of publishers who use them. The first study also addresses RQ2 by assessing the variations in publishers' pay-or-tracking wall designs. The second study delves deeper into RQ2 by examining publishers' motivations for pricing the pay option. It aims to determine whether publishers align the pay option's price with the revenue generated by the tracking option.

The third study addresses RQ3 by investigating how many users leave a publisher when it introduces a pay-or-tracking wall. To do so, we studied multiple publishers' online traffic following their introduction of a pay-or-tracking wall. Finally, the fourth study also addresses RQ3 by examining whether users choose the pay or tracking option when deciding to remain and then calculates the share of both options in relation to multiple online traffic measures (e.g., the number of page impressions, visits, and users). Lastly, we combine the insights from the four empirical studies to answer RQ4 and outline the economic consequences of this approach.

*Table 2: Overview of Empirical Studies*

| Study | Research Question | Title | Major Data Source |
|-------|-------------------|-------|-------------------|
| 1 | RQ1 & RQ2 | Popularity and Implementation of Pay-or-Tracking Walls | **Self-Collected Data:**<br>– Data extracted via automated web scraping and manual collection<br>– Includes the Top-50 publishers in 21 European countries |
| 2 | RQ2 | Motivation for the Pricing of the Pay Option | **Ad Price Data:**<br>– Data from a European ad exchange<br>– Random sample of 467,000 cookies and 325 million sold ad impressions<br>– Includes two Top-50 German publishers |
| 3 | RQ3 | Impact of the "Leave" Option on Online Traffic | **Online Traffic Data:**<br>– Daily traffic metrics recorded by the German Audit Bureau of Circulation<br>– Spanning over 1,341 days<br>– Includes nine of the Top-50 German publishers who adopted a pay-or-tracking wall |
| 4 | RQ3 | Share of the "Pay" and "Tracking" Options in Online Traffic | **Clickstream Data:**<br>– Proprietary data from a Top-50 German publisher<br>– Comprising the browsing behavior of millions of daily users over 517 days |

### Study 1 – Popularity and Implementation of Pay-or-Tracking Walls

***Setup of Empirical Study***

The first empirical study explores in three steps: (i) the popularity of pay-or-tracking walls among top European publishers, (ii) the adoption time, and (iii) publishers' implementation.

*Collection of Data on the Popularity*

Our data collection includes 21 European countries that implemented the GDPR and the 50 most-visited publishers for each country. The list of publishers builds upon Similarweb's website ranking from September 2022; the data collection took place between October 31 and November 3, 2022. We used an automated Python script to screenshot the websites' starting pages. Subsequently, we manually annotated whether the publishers displayed a pay-or-tracking wall. In cases of unclear classification, we manually visited the publisher's website for a more detailed inspection. Further, we categorized the publishers that used a pay-or-tracking wall (e.g., news, weather forecast, search engine, social media).

*Collection of Data about the Adoption Time*

For the publishers that used a pay-or-tracking wall, we determined the adoption time based on the Wayback Machine of the Internet Archive. The Internet Archive provides historical prints of websites. While it typically cannot show the implemented notification banner to obtain consent for tracking, it does allow us to track the publishers' privacy policy updates. Specifically, we assigned the date of introduction as the day we first observed the privacy policy describing the pay-or-tracking wall. Granted, the Internet Archive sometimes provides too few historical prints of the privacy policy. In such cases, we searched for other sources, such as articles introducing the publisher's users to the pay-or-tracking wall or social media posts of users reporting the introduction. In almost all of the cases, we were able to determine the specific day of the introduction.

*Collection of Data about Publishers' Implementation*

We collected data about the publishers' implementation of their pay-or-tracking walls based on the design choices outlined in Table 1. Typically, the descriptions on the publishers' websites include all the necessary information to analyze the differences between the pay and tracking options for advertising, content, and price. Additionally, we bought the pay option to be able to measure its tracker usage. To detect third-party tracking, we used the browser extension Ghostery Insights, which logs network traffic, third-party requests and filters for trackers by comparing them to known tracking domains. We visited each publisher's landing page and two other web pages to record tracker usage—once using the tracking option and once with the pay option. Note that our approach can only detect third-party tracking and not first-party tracking. The latter is often closely integrated with technical website functionalities and thus difficult to distinguish from technical functionalities (due to sharing the same domain and resources).

### Results of Empirical Study

*Popularity of Pay-or-Tracking Walls*

As of November 2022, pay-or-tracking walls were in use among top-50 publishers in Austria, France, Germany, and Italy. We did not observe pay-or-tracking walls among the top-50 publishers in the other 17 European countries we analyzed[2]. The share of pay-or-tracking walls among the country's top-50 publishers was highest in France (20%), followed by Germany (18%), and Austria and Italy (10% each) (see Figure 3). These pay-or-tracking walls originated from 26 different publishers, including three listed among the top-50 publishers of more than one country (e.g., German-speaking audiences in Austria and Germany). Most of these publishers (65%) primarily provide news, while the rest consist of lifestyle magazines and websites on cooking, gaming, sports, TV guides, and weather forecasts. All publishers offer content for users.

---

[2] The other 17 European countries included Belgium, Bulgaria, Croatia, Czechia, Denmark, Finland, Greece, Hungary, Ireland, Netherlands, Poland, Portugal, Romania, Slovakia, Spain, Sweden, and United Kingdom.

*Figure 3: Share of Pay-or-Tracking Walls among Top-50 Publishers per European Country*

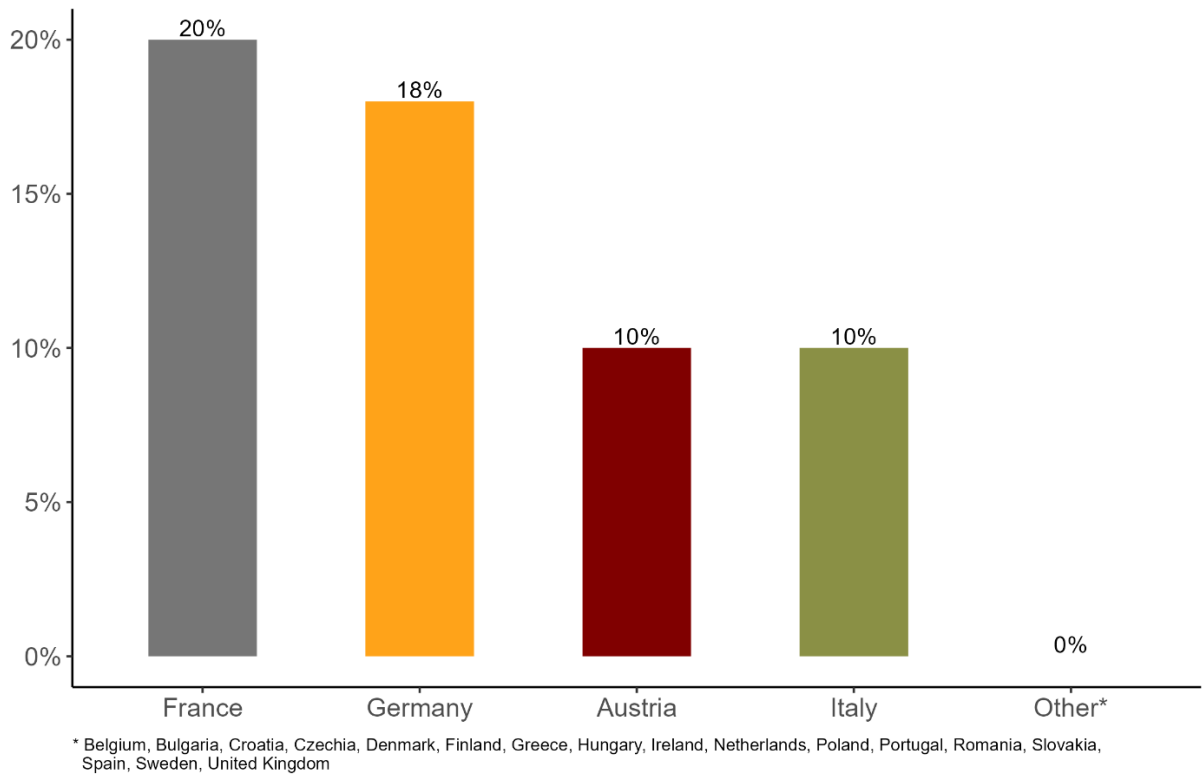

\* Belgium, Bulgaria, Croatia, Czechia, Denmark, Finland, Greece, Hungary, Ireland, Netherlands, Poland, Portugal, Romania, Slovakia, Spain, Sweden, United Kingdom

*Distribution of Adoption Time*

As depicted in Figure 4, the Austrian news website "Der Standard" pioneered the introduction of pay-or-tracking walls in 2018. In 2020, the first two German publishers and another one in Austria followed. However, the practice really took off after April 2021 with the entrance of French publishers. The adoption in France coincided with the end of the grace period for the French Data Protection Authority's new guidelines for cookies (French Data Protection Authority 2020). This guideline appears to have spurred immediate implementation at eight publishers, four of which belong to one legal entity. Publishers in Italy joined the ranks starting in 2022.

*Figure 4: Share of Pay-or-Tracking Walls among Top-50 Publishers per Country over Time*

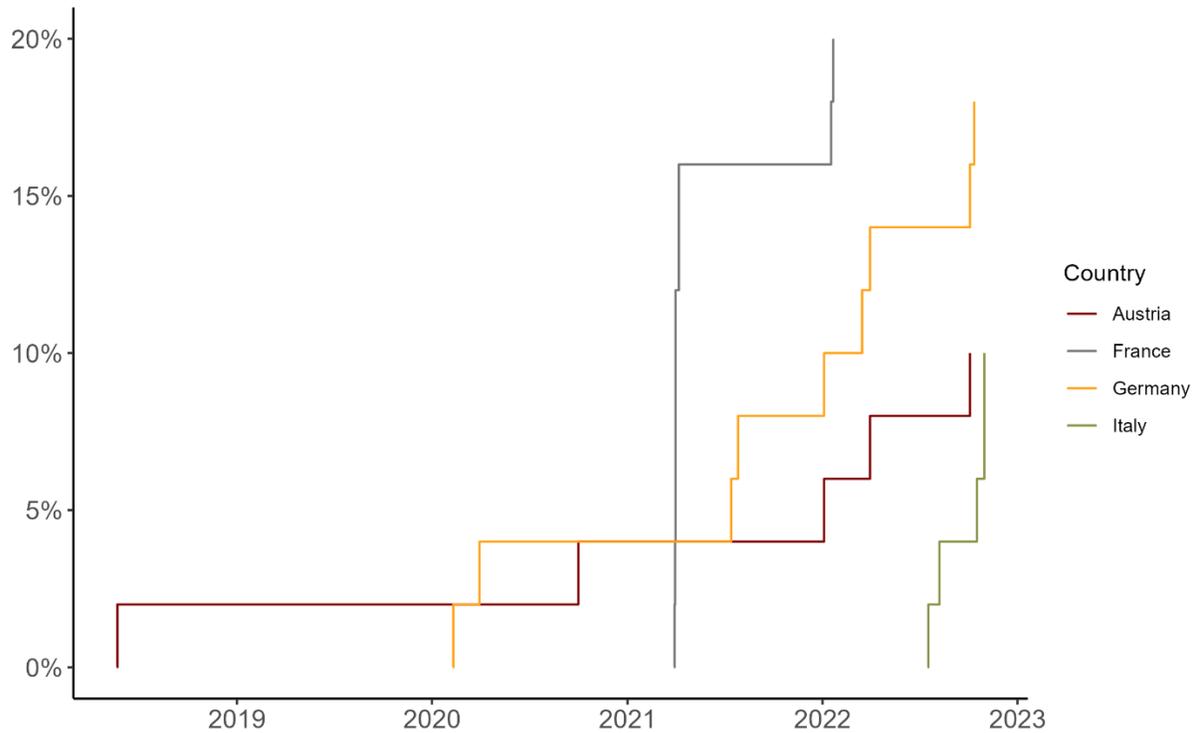

*Publishers' Implementations of Pay-or-Tracking Walls*

As Table 3 outlines, the publishers exhibited considerable variation in their implementations, mainly due to the diverse use of bundling strategies related to paid content offerings. Some variation stemmed from having a price spectrum for the pay option alongside mixed usage of bundling the pay option with advertising-free access. We observed fewer differences regarding tracking and other price-related aspects, such as the consistent use of flat rate models for the pay option. We subsequently elaborate on the details.

*Table 3: Overview of Publishers' Implementations of Pay-or-Tracking Walls*

| Dimension | Feature | Tracking Option | Pay Option |
|---|---|---|---|
| **Tracking** | *Number of* Trackers | [25; 104] | [0; 18] |
| | *Avg. Reduction of* Trackers (vs. Tracking Option) | N/A | -87% |
| | *Share of Publishers* without Trackers ("Tracking-Free") | 0% | 4% |
| **Advertising** | *Share of Publishers* without Advertising ("Advertising-Free") | 0% | 58% |
| **Content** | *Share of Publishers* with Paid Content Offering | 62% | |
| | *If Paid Content Offering: Share of Publishers with* | | |
| | a) Price bundling | 63% | |
| | b) Product bundling | 31% | |
| | c) No bundling | 6% | |
| **Price** | *Price of* Cheapest Available Option (in EUR per Month)[a] | 0.00 | [0.41; 11.74] |
| | *Share of Publishers with* | | |
| | Minimum Commitment Period of a Month | N/A | 85% |
| | Automatic Renewal Policy | N/A | 69% |
| | (Partially) Anonymous Payment Methods | N/A | 0% |
| | Requirement to Create Account with Personal Details | 0% | 88% |
| **Other** | *Share of Publishers* Participating in Joint Pay Option | N/A | 19% |

[a] Price based on the minimum commitment period, without promotional deals and exclusive value-added taxes.

*Notes:* N = 26 Publishers, N/A = not applicable

## Description of the Amount of Tracking

The upper part of Table 3 indicates that publishers integrated between 25 and 104 third-party trackers in their tracking options. In the pay option, they reduced the number of trackers by 87%, on average. That said, publishers did typically still implement third-party trackers into the pay option. Only one publisher had a pay option that was tracking-free (4% of the analyzed publishers). In Web Appendix A, we categorize the third-party trackers used in the pay option. We show that the trackers in the pay option are not only used for one specific functionality. Instead, the publishers used different trackers from multiple categories, such as advertising, site analytics, or social media.

*Bundling Pay-or-Tracking Walls with Advertising-Free Access*

Across all analyzed pay-or-tracking walls, the tracking option came with advertising, allowing publishers to monetize their ad inventory via behavioral targeting. Meanwhile, 42% of the publishers displayed advertising to paying users. More publishers (58%) instead utilized a bundling strategy (see Table 3), meaning that paying users no longer saw ads on the entire website and app, save for a few exceptions where ads still occur on certain parts.

*Bundling Pay-or-Tracking Walls with Additional Content*

Thirty-eight percent of the analyzed publishers had no paid content offering; most of these were non-news publishers. These publishers enabled unlimited access to their content to both users of the pay and tracking options. As outlined in Table 3, most publishers (62%) used paid content and commonly bundled it with the pay-or-tracking wall. About a third of those publishers using a bundling strategy employed price bundling (i.e., there were discounts for purchasing both the pay option and access to unlimited content). The other two-thirds engaged in product bundling, wherein users mostly could only receive the pay option and access to unlimited content as a package.

In summary, we observed considerable variation between the implementations (Table 3): from publishers who offered no paid content at all, to those who bundled paid content, to one exceptional publisher who offered paid content without any form of bundling.

*Distribution of Price Models and Prices*

Even when using a paid content strategy, all the examined publishers offered at least one costless tracking option. As outlined in Figure 5 (and summarized in Table 3), the prices of the publishers' cheapest available pay options varied between 0.41 EUR and 11.74 EUR per month (excluding value-added taxes). The pay option always involved a flat rate with a fixed price, reflecting a pay-per-period model. We did not observe other price models, such as the pay-per-use model with usage-dependent pricing. As depicted in Table 3, most pay options (85%) included a minimum commitment period of a month.

Moreover, 69% of the pay options were subscription-like and renewed automatically, compared to 31% canceling automatically after the commitment period (see Table 3). The publishers enabled common payment methods, including bank transfers, credit cards, or payment service providers like PayPal. However, we did not observe any (partially) anonymous payment methods that might allow users to hide their payment details from the publisher or payment service provider. Additionally, in 88% of the cases, publishers asked paying users to create an account, which at least requires the user's email address (see Table 3). Only 12% of the publishers used alternative authentication methods to control access to the pay option, such as by providing an identification code that the user had to enter when initially visiting the publisher's website or app.

*Usage of a Joint Pay Option*

Notably, 19% of the examined publishers participated in a joint pay option, wherein one pay option granted access to multiple publishers and reduced tracking across all their websites and apps. (see Table 3). One of these joint pay options involved a group of French publishers belonging to the same legal entity, while another was a collaborative offering of independent German publishers.

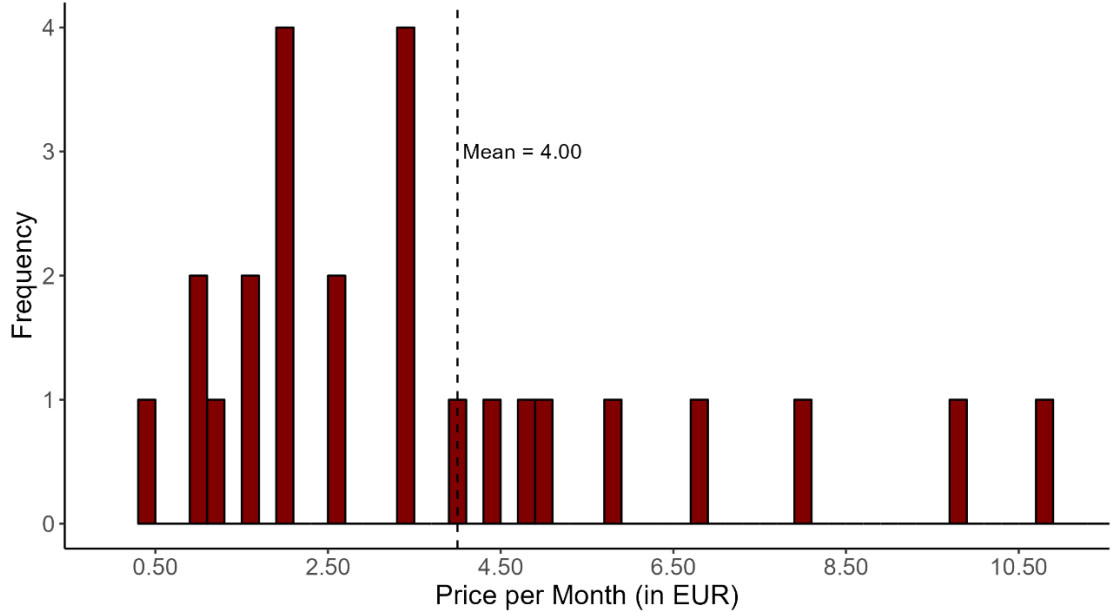

*Figure 5: Prices of the Pay Option*

Notes: Prices based on the minimum commitment period, without promotional deals and exclusive value-added taxes. Prices refer to the publishers' cheapest available pay option in case multiple options were offered, such as when bundling with paid content is used.

## Study 2 – Motivation for the Pricing of the Pay Option

### *Setup of Empirical Study*

In this study, we gain further insights into how publishers implement pay-or-tracking walls by examining publishers' motivations for pricing the pay option. Specifically, we investigate whether publishers set the pay option's price so that it compensated for the alternatively generated revenue if the user decides on the tracking option, i.e., the ad revenue when tracking is possible.

To this end, we first had to consider whether publishers displayed advertising under the pay option or not. If not, then we compared the pay option's price with the "foregone ad revenue". If so, then we compared the pay option's price with the "decrease in ad revenue due to the inability to track the user".

Scholars have achieved mixed results on the exact value added by tracking. Four studies have addressed this question directly, varying in their sample of publishers and their dependent ad price variable. The first one, by Marotta, Abhishek, and Acquisti (2019), investigated ad prices that one publisher received and found an average ad price decrease of 8% without ad tracking. The second one, by Wang, Jiang, and Yang (2024), examined ad prices that one publisher receives and observed a 5.7% decrease without ad tracking. The third one, by Johnson, Shriver, and Du (2020), analyzed the ad prices paid by advertisers based on data from an ad exchange with multiple publishers, finding an average decrease of 52% without ad tracking. The fourth one, by Laub, Miller, and Skiera (2023), analyzed ad prices that publishers received based on data from an ad exchange with multiple publishers and found an average decrease of 18% without ad tracking. Since the latter article's setting most closely matched our own (i.e., the authors base their analysis on multiple publishers and the ad prices publishers receive, which directly correspond to publisher revenues), we assumed an average ad price decrease of 18% for our calculations when tracking was not possible.

*Grouping of Publishers According to their Pay Option's Prices*

For the 26 publishers that used a pay-or-tracking wall, as depicted in Figure 5, we differentiated between those displaying advertising under the pay option and those that did not.

*Description of Ad Price Data*

We used data from a European ad exchange that reaches 84% of Internet users in its respective market. The data comprise a random sample of 467,000 cookies and 325 million ad impressions sold between 2014 and 2016. It only includes ad impressions with a cookie identifier, thus, considering ad impressions when tracking is possible. Each ad impression includes the publisher, a timestamp, and the price paid to the publisher.

*Calculation of the Ad Revenue per User*

As the pay option's prices comprise only the top-50 publishers in their respective countries, we similarly filter the ad price data for the top-50 publishers. This filtering resulted in a sample of two German top-50 publishers that were available in the ad price data. To derive the distribution of ad revenue that a publisher earns for a user ("cookie") per month, we summed up the prices paid for the cookie's ad impressions at a particular publisher and in the corresponding month. We also accounted for incomplete months of our data: For instance, we might observe a cookie for 1.5 months, in which case we can reasonably assume that the data include all displayed ads for the first month. However, for the second month, we cannot say whether the user stopped accessing the publisher's content or if the cookie was deleted and replaced by a new cookie. Therefore, we excluded incomplete months (i.e., less than an entire month) from our data.

Further, we only considered users who seriously engaged with the publisher's content by filtering for cookies showing engagement for at least five days within a month with the publisher's content. Our final sample consisted of 172,082,120 ad impressions from two publishers and 77,681 cookies observed over about 2.5 years, reflecting 374,322 observations for the monthly ad revenue per user and publisher. Each observation included, on average, 459.7 ad impressions.

*Calculation of the Decrease in Ad Revenue Due to the Inability to Track Users*

Following the results of Laub, Miller, and Skiera (2023), we used an 18% decrease in ad revenue as the baseline decrease when tracking was not possible.

### Results of Empirical Study

*Distribution of the Ad Revenue per User*

As depicted in Figure 6, we estimated that a top-50 publisher, on average, generates monthly ad revenue of 0.24 EUR per user when tracking is possible. We further observed a long-tail distribution, meaning that a small share of users generates larger ad revenues. In the 99% percentile, for instance, users generate more than 2.01 EUR.

*Figure 6: Monthly Ad Revenue per User when Tracking is Possible*

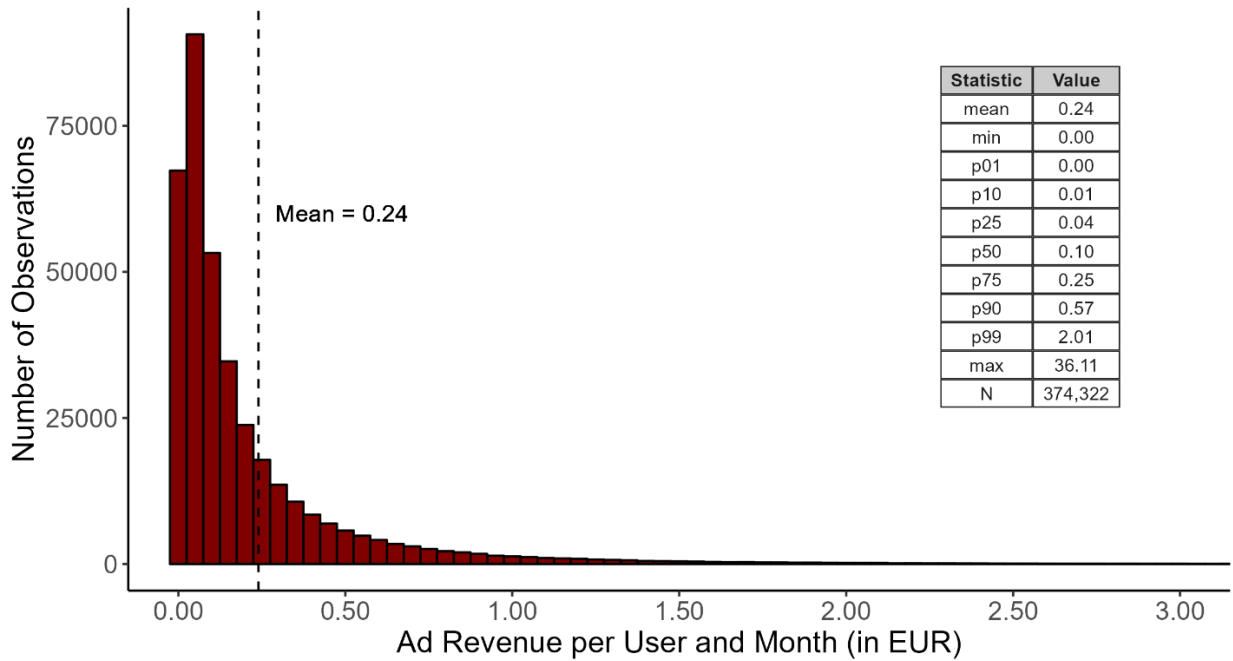

Notes: min = minimum, p01 = 1% percentile, p10 = 10% percentile, p25 = 25% percentile, p50 = 50% percentile, p75 = 75% percentile, p90 = 90% percentile, p99 = 99% percentile, max = maximum, N = number of observations. An observation refers to the combination of a month and user. For example, observing 10 users and each user for 3 months yields 30 observations.

*Comparison of Pay Option's Prices With and Without Advertising*

In Figure 7, we show the distributions of the pay option's price: one for publishers that display advertising under the pay option and one for those that do not. Both distributions exhibit a similar degree of variation, with no pattern indicating that the prices are higher for pay options without advertising than those with advertising. Conversely, the average price for pay options without (vs. with) advertising is 3.24 (vs. 5.04) EUR.

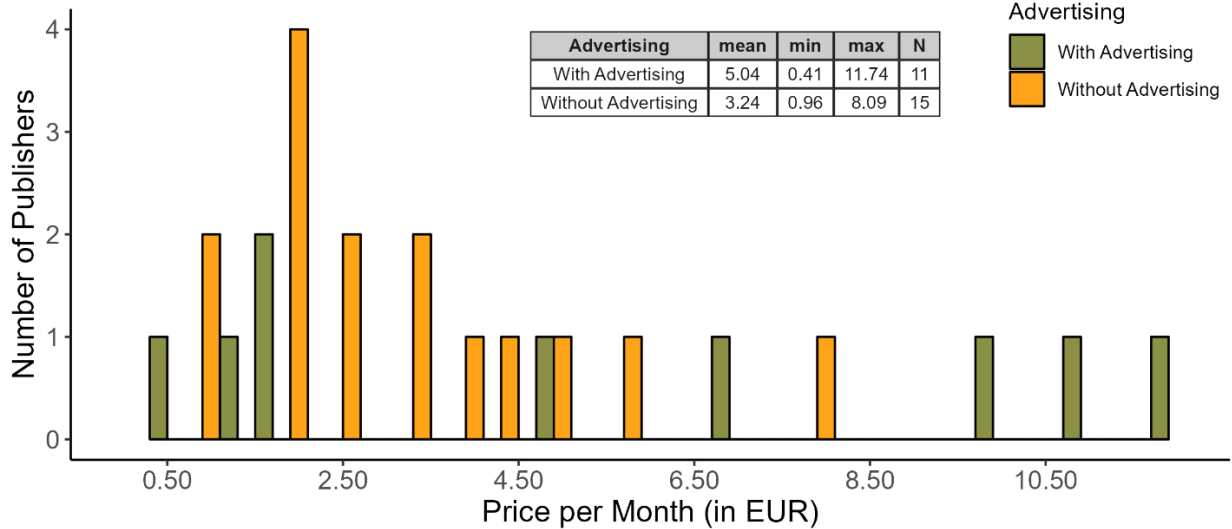

*Figure 7: Pay Option's Prices with and without Advertising*

| Advertising | mean | min | max | N |
|---|---|---|---|---|
| With Advertising | 5.04 | 0.41 | 11.74 | 11 |
| Without Advertising | 3.24 | 0.96 | 8.09 | 15 |

Notes: Prices based on the minimum commitment period, without promotional deals and exclusive value-added taxes.
Prices refer to the publishers' cheapest available pay option. Explanations: min = minimum, max = maximum,
N = number of publishers.

*Comparison of the Pay Option's Price to the Ad Revenue When Tracking Is Possible*

In the upper panel of Figure 8, we compare the pay option's prices without advertising (orange bars in Figure 7) with the foregone ad revenue (Figure 6). We observe that the prices coincide with the upper end of the ad revenue distribution. For instance, the average price of the pay option of 3.24 EUR corresponds with the 99.65% percentile of the ad revenue distribution.

In the lower panel of Figure 8, we compare the pay option's prices with advertising (green bars in Figure 7) with the decrease in ad revenue arising from the inability to track the user (18% times the values of Figure 6). Similarly, we observe that the prices are typically above the decrease in ad revenue. For instance, the average price of the pay option of 5.04 EUR corresponds with the upper 99.99% percentile of the ad revenue decrease distribution.

Thus, our results indicate that publishers do not base their prices for the pay option on the alternative ad revenue of the tracking option.

*Figure 8: Comparing the Pay Option's Prices to the Ad Revenue When Tracking Is Possible*

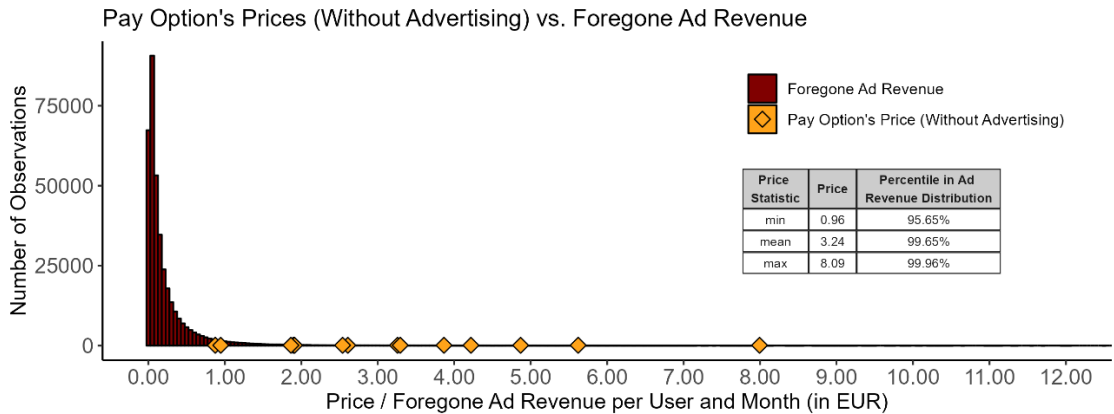

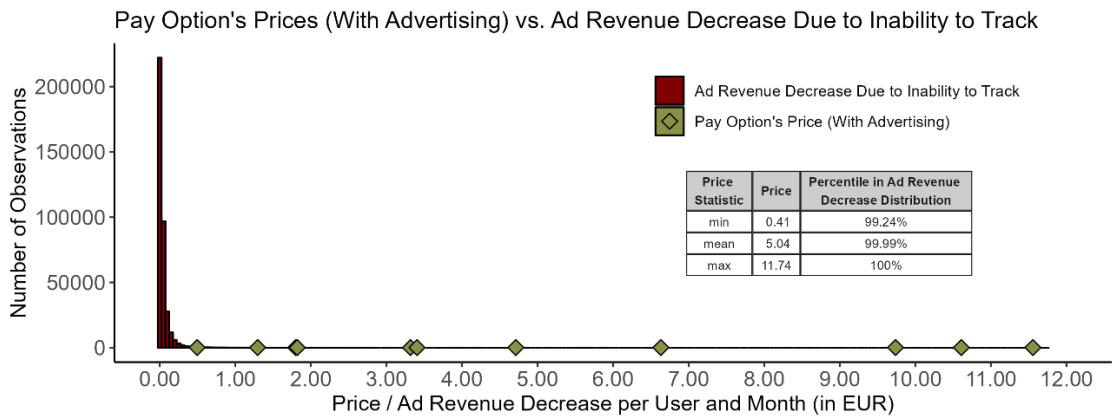

Notes: Ad Revenue Decrease Due to Inability to Track = 18%-Share of Ad Revenue per User. An observation refers to the combination of a month and user. For example, observing 10 users and each user for 3 months yields 30 observations.

## Study 3 – Impact on Online Traffic ("Leave Option")

### Setup of Empirical Study

In the third empirical study, we examine whether the publishers' online traffic decreases after implementing a pay-or-tracking wall. Users who accessed a publisher's content before might decide to leave and abandon the publisher after being confronted with the pay-or-tracking wall. Similarly, new users might not engage with the publisher's content due to the pay-or-tracking wall. Ultimately, such user behavior may impact the publisher's online traffic, potentially resulting in fewer page impressions and visits.

*Description of Online Traffic Data*

We obtained access to the daily online traffic of over a thousand German publishers, spanning 1,341 days between May 2019 and December 2022. These data include the German top-50 publishers, which we had previously identified as adopters of a pay-or-tracking wall. The traffic data derive from the German Audit Bureau of Circulation, which records publishers' online traffic to provide comparable and reliable performance information to the advertising industry. Participating publishers integrate a standardized script that counts the number of daily page impressions across their offerings, including their website and app. Publishers conduct the anonymized measurement of the number of page impressions independently of the users' consent to tracking, such as for their pay-or-tracking wall. Note, however, that the German Audit Bureau of Circulation provides an opt-out mechanism through its website, which requires users to store a specific opt-out cookie. Aside from the page impressions, the data include the daily number of visits until November 2021 based on a purely technical measurement.[3]

*Description of the Adoption Times of Pay-or-Tracking Walls*

We reused the data from Study 1 about the top-50 publishers' initial date of adopting their pay-or-tracking walls. The data provide us with precise information regarding the dates on which publishers introduced their pay-or-tracking walls.

*Description of Google Trends Data on News Interest*

We used data from Google Trends to explore and analyze the popularity of search terms on the Google search engine. Specifically, we extracted information about users' search interest for the search term "news" (in German: "Nachrichten") within Germany between May 2019 and

---

[3] From December 2021, due to a change in the legal environment, the reported number of visits is an estimate based on the visits of users who consented to tracking. Therefore, we do not consider the number of visits after November 2021.

December 2022. This dataset serves as a proxy for capturing users' interest in news content over time. The data came as a weekly index between 0 and 100, where a higher index value reflects a greater search interest.

*Description of the Dependent Variable for Online Traffic*

The number of daily page impressions served as our main dependent variable. Its technical and anonymized measurement allowed us to analyze the change in traffic based on all users and independently of the users' consent decision to their tracking. Other potentially insightful traffic metrics, such as the bounce rate (i.e., the share of users leaving a website just after one page impression), typically require a non-anonymized measurement that connects a user identifier with the number of page impressions, but that goes beyond the data provider's aggregated and anonymized traffic metrics.

*Description of Sample*

Our analysis focused on the nine top-50 German publishers that we had previously identified as using a pay-or-tracking wall. We excluded one publisher from the sample due to missing data around its introduction of the pay-or-tracking wall. We also excluded another publisher that provides weather forecasts as its core service, since it likely follows different time trends as opposed to the remaining news publishers. Our final sample thus consisted of seven top-50 German publishers who primarily offer national news to their users.

*Description of Approach with Before-After Comparison*

Our primary approach involved a before-after comparison, where we compared each publisher's traffic before and after the introduction of its pay-or-tracking wall. This approach utilizes the before-period as the control group for the comparison, which allows us to eliminate the possibility of spillover effects. The latter could affect our results if we used a different control group, such as publishers without a pay-or-tracking wall, potentially leading to an estimation bias.

Our methodology's primary challenge is avoiding any other events occurring in the before- and after-period (Huntington-Klein 2021). Therefore, we chose a short time window: six weeks before vs. six weeks after the introduction of the pay-or-tracking wall. This short time window reduces the influence of other considerable events, while likely being long enough to detect a possible change in online traffic. We also considered control variables, such as users' search interest for the search term "news" over time. This consideration allowed us to capture the influence of important news events, such as the pandemic, that temporarily boosted users' interest in news content.

*Description of Limitations and Robustness Tests*

While we aimed to minimize the impact of other concurring events, we cannot rule out the possibility of an estimation bias due to our specific choice of a time window. Moreover, our primary dependent variable, the daily number of page impressions, might be susceptible to the effects of temporal trends. Therefore, we performed a robustness test and repeated our analysis using various time windows and alternative dependent variables, such as the daily number of visits or the number of page impressions per visit (see Web Appendix B).

Even though we considered the impact of other events and time trends via control variables, we cannot completely rule out their possible impact. Thus, we used a different identification strategy in another robustness test (see Web Appendix C). More specifically, we applied a synthetic difference-in-difference approach, following Arkhangelsky et al. (2021). In this approach, we utilized publishers' different adoption times to create a synthetic control group for each publisher.

Lastly, users might anticipate the introduction of a publisher's pay-or-tracking wall and abandon said publisher preemptively. However, we found no evidence of publishers communicating their change to users before introducing their pay-or-tracking wall.

### Results of Empirical Study

*Graphical Analysis of Online Traffic*

We first present some model-free graphical analysis. As depicted in Figure 9, we used the weekly number of page impressions to create a weekly index for each publisher. We normalized each index to a baseline value of 100 in the week preceding the introduction of the pay-or-tracking wall. This normalization allowed us to examine the online traffic patterns of the publishers collectively, despite variations in the volume of their page impressions.

As illustrated in Figure 9, most publishers exhibited weekly page impression indices that centered around the baseline value of 100; some experienced an increase following the implementation of their pay-or-tracking wall. Conversely, two publishers seemed to have considerably fewer page impressions after the introduction, but also showed a spike in the weeks before the introduction of their pay-or-tracking wall. A closer examination revealed that the two publishers were affected by two major news events: one being the beginning of the pandemic in March 2020 and the other being the start of the Ukraine war in February 2022. In summary, the model-free graphical analysis suggests that most publishers' online traffic did not considerably decline when introducing a pay-or-tracking wall.

*Figure 9: Weekly Page Impression Index per Publisher over Time*

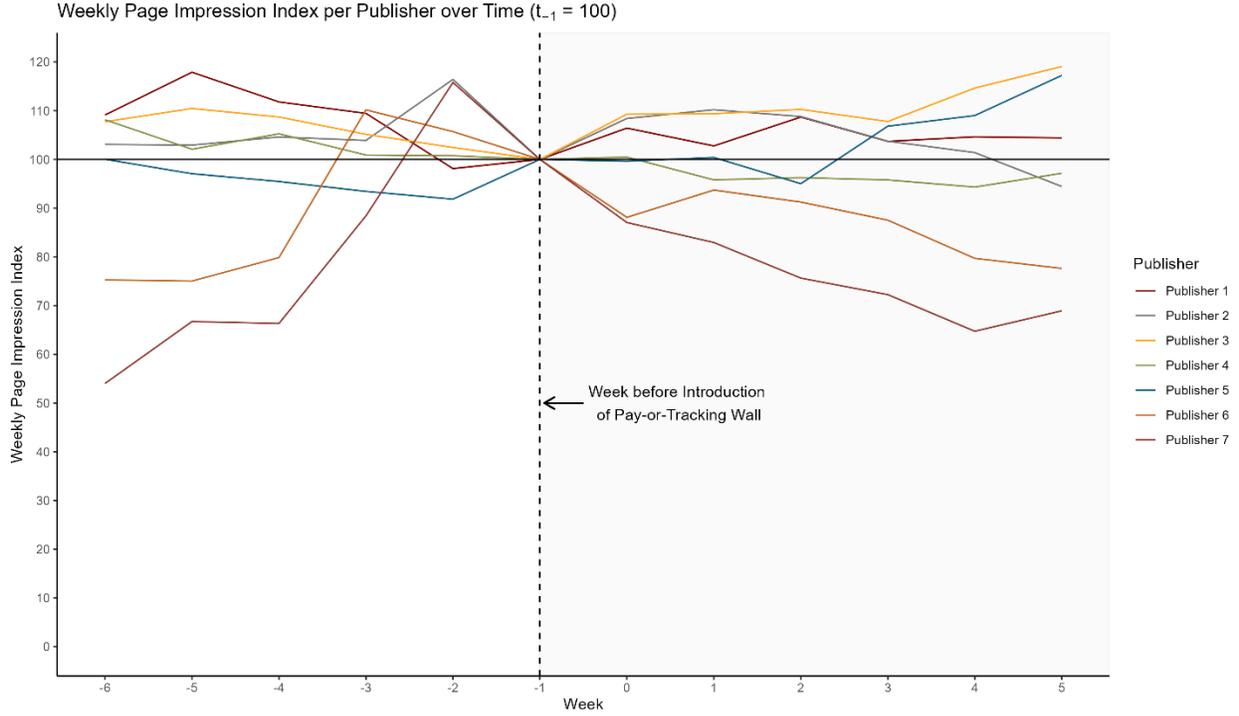

*Impact of the Introduction of a Pay-or-Tracking Wall on Online Traffic*

We analyzed the impact of introducing a pay-or-tracking wall following our primary approach of a before-after comparison. We used an ordinary least squares (OLS) regression to obtain the effect's confidence intervals and included control variables for time trends. Further, we applied a log transformation to our dependent variable, the number of daily page impressions ($\ln(\#\ Daily\ Page\ Impressions)$), since we were interested in percentage changes. We estimated three distinct models, which together allowed us to evaluate the impact of the pay-or-tracking wall while controlling for various factors, including publisher ($i$) fixed effects, day-of-week ($d$) fixed effects, and the variation among users' general news interest. The specifications for our OLS models are as follows:

$$\ln(\#\ Daily\ Page\ Impressions_{it}) = \alpha_i + \beta After_t + \epsilon_{it} \quad (1)$$

$$\ln(\#\ Daily\ Page\ Impressions_{itd}) = \alpha_{id} + \beta After_t + \epsilon_{it} \quad (2)$$

$$\ln(\# \ Daily \ Page \ Impressions_{itd}) = \alpha_{id} + \beta After_t + \gamma NewsInterest_{it} + \epsilon_{it} \quad (3)$$

The first model incorporated fixed effects for each publisher ($\alpha_i$) to account for variations in traffic volume among each publisher $i$. The coefficient of interest, $\beta$, is based on an indicator variable ($After_t$), which takes the value 1 for all days $t$ after introducing the pay-or-tracking wall. In the second model, we introduced fixed effects ($\alpha_{id}$) for each publisher $i$ and the day of the week $d$ to consider the influence of weekday patterns. In the third model, we further accounted for variations among users' general news interest by including the Google Trends coefficient and variable ($\gamma NewsInterest_{it}$), which measures users' interest in the search term "news". This variable is publisher-specific since each publisher introduced its pay-or-tracking wall on different dates, which impacts index $t$. It serves as a control variable to capture the impact of important news events, as illustrated in Figure 10.

*Figure 10: Google Search Trend in Germany over Time for the Search Term "News"*

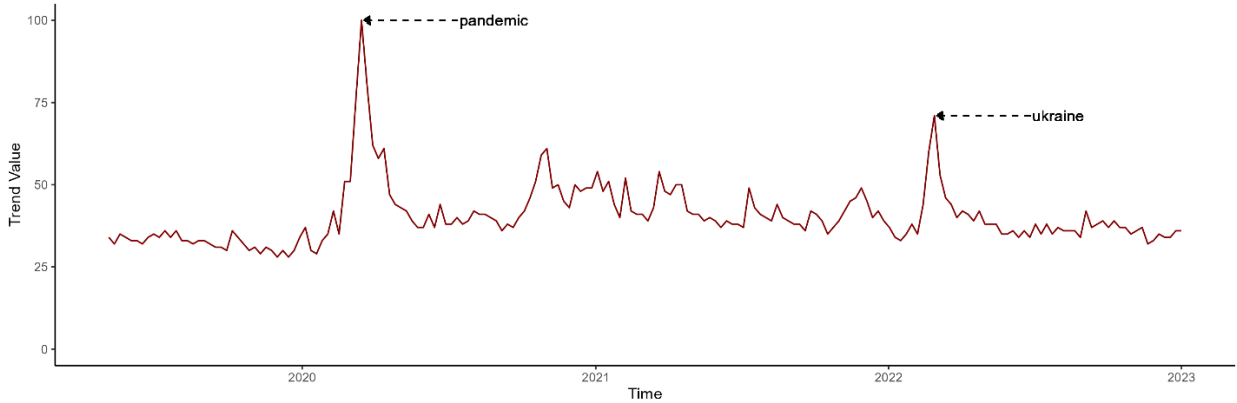

We present the results of our estimation in Table 4. The coefficient of interest, $\beta$, representing the change of online traffic in the after-period, is not statistically different from zero in all three models. This indicates that publishers did not experience a decline in their online traffic after implementing a pay-or-tracking wall. The robustness tests confirm the result when using various time windows and alternative dependent variables (see Web Appendix B) or a different identification strategy via the synthetic difference-in-difference estimator (see Web Appendix C).

*Short-Term Impact of the Introduction of a Pay-or-Tracking Wall on Online Traffic*

Next, we modified our estimation to examine whether introducing a pay-or-tracking wall exerted a short-term effect. Instead of estimating a single coefficient for the after-period, we included a time-varying coefficient ($\beta_w$) for each included week $w$, with the week preceding the introduction of the pay-or-tracking wall serving as the reference. In this way, we could observe separate estimates for each week over time for all three models:

$$\ln(\# \, Daily \, Page \, Impressions_{itw}) = \alpha_i + \sum_{w=-6}^{W=5} \beta_w Week_w + \epsilon_{it} \quad (4)$$

$$\ln(\# \, Daily \, Page \, Impressions_{itdw}) = \alpha_{id} + \sum_{w=-6}^{W=5} \beta_w Week_w + \epsilon_{it} \quad (5)$$

$$\ln(\# \, Daily \, Page \, Impressions_{itdw}) = \alpha_{id} + \sum_{w=-6}^{W=5} \beta_w Week_w + \gamma NewsInterest_{it} + \epsilon_{it} \quad (6)$$

Figure 11 reveals the short-term effects of implementing a pay-or-tracking wall on online traffic. The weekly coefficients were not statistically different from zero in all three models (4), (5), and (6), suggesting that publishers did not experience considerable short-term effects when introducing a pay-or-tracking wall.

Overall, our analysis suggests that implementing a pay-or-tracking wall leads to no significant decrease in publishers' online traffic. Users do not seem to leave and abandon the publisher after being confronted with a pay-or-tracking wall.

*Table 4: Results of the Before-After Comparison of the Publishers' Online Traffic*

| Dependent Variable: | ln(Daily Number of Page Impressions) | | |
|---|---|---|---|
| Model: | (1) | (2) | (3) |
| After | -0.0066 (0.0213) | -0.0066 (0.0213) | 0.0376 (0.0296) |
| News Interest | | | 0.0102*** (0.0012) |
| Publisher Fixed Effects | ✓ | | |
| Publisher-Specific Day-of-Week Fixed Effects | | ✓ | ✓ |
| N Observations | 588 | 588 | 588 |
| $R^2$ | 0.9790 | 0.9898 | 0.9948 |
| Within $R^2$ | 0.0004 | 0.0008 | 0.4855 |

Significance levels: * $p < 0.05$, ** $p < 0.01$, *** $p < 0.001$.
Standard errors are clustered at the publisher level and reported in parentheses.
Notes: This table shows the coefficient (After) from the OLS regressions of the before-after comparison. We include the six weeks before and after the introduction of each publisher's pay-or-tracking wall. Multiplying the number of publishers (N publishers = 7) and the number of days of the included 12 weeks (T = 84 days) yields the number of observations (N observations = 588).

*Figure 11: Development of the (Weekly) Coefficient for Online Traffic*

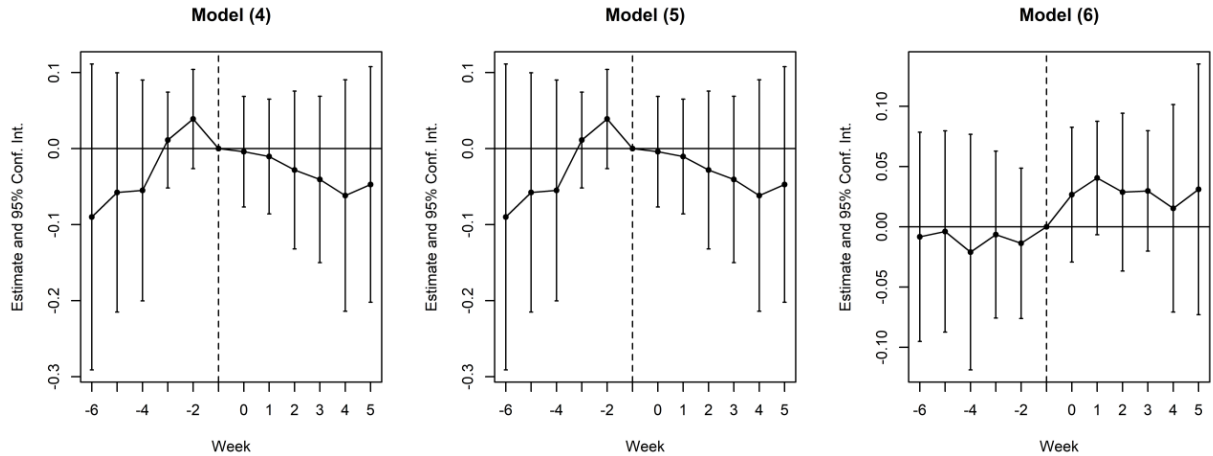

Notes: The dashed vertical lines denote the introduction of the pay-or-tracking wall.

## Study 4 – Share of the "Pay" and "Tracking" Options in Online Traffic

### Setup of Empirical Study

The fourth empirical study examines whether users choose the pay-or-tracking option (conditional on expressing a choice for one of the options). To this end, we utilized the proprietary clickstream data of a publisher that ranks among the top-50 publishers and the top ten national news outlets in Germany with millions of daily users.

This publisher presents its pay-or-tracking wall as part of a paid bundle with advertising-free access. Moreover, the publisher offers two access plans for paid content: a limited and an unlimited access plan. Users face no charge if they choose the tracking option and the limited access plan. A charge occurs for all other combinations (pay option + limited access, pay option + unlimited access, tracking option + unlimited access). The pay option's price with the limited access plan is about 4.25 EUR per month (excluding value-added taxes). As the publisher engages in a price bundling strategy, users pay about 50% less for the pay option if they also pay for the unlimited access plan.

*Description of Clickstream Data*

Thanks to the publisher's granular data, we can calculate multiple traffic measures and distinguish between users choosing to pay or be tracked. The data begin 52 days after the introduction of the pay-or-tracking wall and cover 517 days (17 months), six days of which we had to exclude due to missing data. The clickstream data encompasses the browsing behavior of millions of daily users—and thus entails billions of observations.

*Calculation of the Share of the Pay and Tracking Options*

We calculated three traffic metrics—number of page impressions, visits, and unique users—on a daily level for the traffic that expressed a choice for either the pay or tracking options. As the publisher prefers to stay anonymous, we only report the share of the pay and tracking options instead of absolute numbers.

**Results of Empirical Study**

Figure 12 depicts the tracking option's share over time for our three metrics (i.e., page impressions, visits, and unique users) among the online traffic involving an expressed choice. While the tracking option's share among the traffic measures decreased over time, it remained predominant compared to the pay option. For instance, on the last observed day, one year and five months after the publisher introduced its pay-or-tracking wall, the tracking option accounted for 99.06% of unique users, 98.65% of visits, and 97.78% of page impressions. Conversely, the pay option accounted for 0.94% of unique users, 1.35% of visits, and 2.22% of page impressions. Clearly, the majority of users choose the tracking option, although paying users generate more page impressions per day than users of the tracking option (14.2 versus 5.8).

*Figure 12: Share of the Tracking Option*
*(as Measured by Page Impressions, Visits, and Unique Users over Time)*

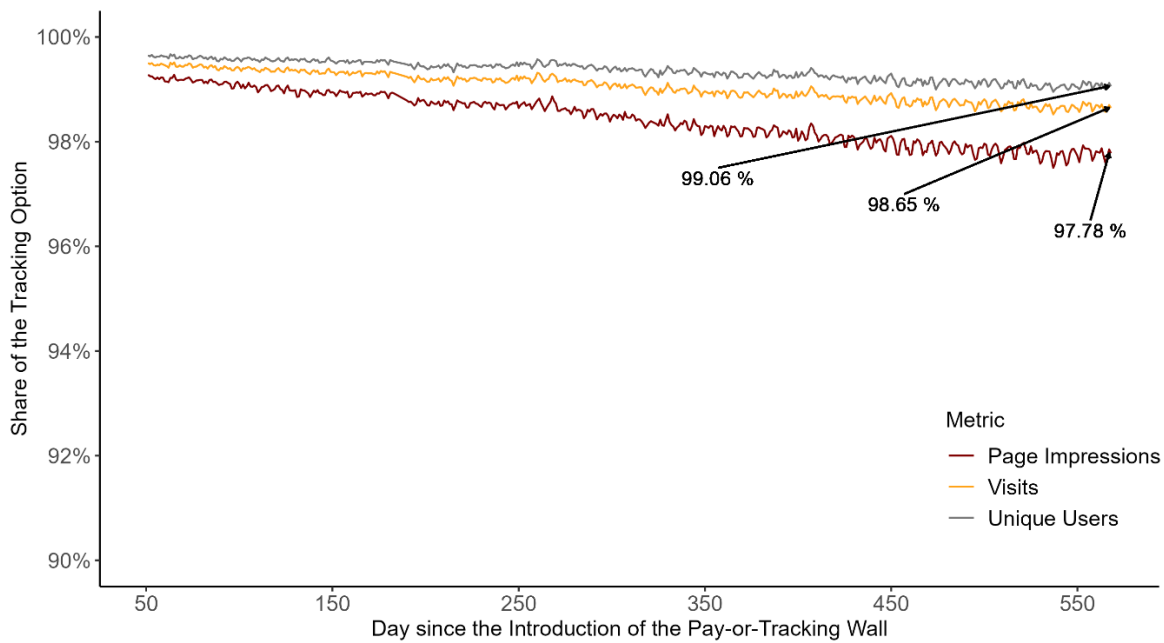

**Economic Consequences of Pay-or-Tracking Walls**

Here, we assess the economic consequences of introducing a pay-or-tracking wall relative to its alternative: a cookie consent banner. Such banners similarly ask users for their consent to being tracked ("tracking option"), but typically allow users to refuse tracking at no cost instead of requiring them to pay ("costless refuse option").

Publishers using a cookie consent banner can monetize their offering by displaying ads. They can earn ad revenue for users who either take the tracking option or the costless refuse option. However, the latter users cannot be used for behaviorally targeted ads and thus generate lower ad revenues due to the higher prices advertisers pay for behaviorally targeted ads relying on tracking (e.g., Johnson, Shriver, and Du 2020; Laub, Miller, and Skiera 2023; Marotta, Abhishek, and Acquisti 2019; Wang, Jiang, and Yang 2024). Publishers using a pay-or-tracking wall similarly offer a tracking option and benefit from this option because it enables them to track users, resulting in higher ad revenue. In addition, they monetize the pay option by charging a fee while still potentially drawing advertising revenues if they display ads.

Pay-or-tracking walls can have two economic disadvantages compared to cookie consent banners: The first is a reduction in demand due to users leaving; the second is a risk of lower revenue due to the chosen price. In what follows, we elaborate on the two potential economic disadvantages and use a back-of-the-envelope calculation to illustrate the economic consequences of implementing a pay-or-tracking wall.

*Description of Potential Economic Disadvantages*

Regarding demand, publishers that adopt a pay-or-tracking wall offer no costless refuse option and thus might experience a decline in online traffic. However, our third empirical study showed that publishers implementing a pay-or-tracking wall did not experience a decrease in their online traffic. Thus, we have no evidence to support a reduction in demand.

Regarding price, publishers may face losses if the pay option's price is lower than the revenue publishers earn with a user of the tracking option (or even lower than the revenue publishers earn with a user of a costless refuse option). However, our second empirical study showed that publishers set their pay options' prices to exceed the alternative ad revenues that publishers otherwise earn per user. Thus, publishers' implementations incur no economic disadvantages related to price.

In summary, introducing pay-or-tracking walls has no adverse economic consequences, and publishers seem to benefit instead. Publishers benefit primarily from users choosing the tracking option, which, on average, exceeds the earnings for the costless refuse option. While the design of cookie consent banners can strongly affect the share of consenting users (e.g., Utz et al. 2019), privacy laws such as the GDPR require publishers to provide a single-click refuse option (e.g., French Data Protection Authority 2021), yielding a share of consenting users of approximately 80% according to research (Jha et al. 2023) and between 70% and 80% according to industry sources (Commanders Act 2023). Meanwhile, our results show that approximately 99% of users choose the tracking option when confronted with a pay-or-tracking wall.

*Back-of-the-Envelope Calculation of the Economic Consequences*

In Table 5, we illustrate the impact on a publisher's revenue with a back-of-the-envelope calculation. In the case of the cookie consent banner, we assume that a publisher has one million users and 80% of them select the tracking option. In the case of the pay-or-tracking wall, we assume that the publisher bundles its pay option with an ad-free experience and thus only earns revenue via the price of the pay option. Additionally, we build upon the results of our four empirical studies.

The results in Table 5 suggest that publishers economically benefit from adopting a pay-or-tracking wall when they have a considerable share of users who refuse to be tracked in a cookie consent banner (a notable revenue increase of 16.4% for the pay-or-tracking wall compared to the cookie consent banner banner). Overall, pay-or-tracking walls seem to be the dominant strategy compared to using a cookie consent banner.

*Table 5: Exemplary Back-of-the-Envelope Calculation of the Economic Consequences*

| | Option | Share of Users | Number of Users[a] | Revenue per User (in EUR) | Revenue per Option (in EUR) | Total Revenue (in EUR) |
|---|---|---|---|---|---|---|
| **Cookie Consent Banner** | Tracking | 80 %[b] | 800,000 | 0.24[d] | 192,000 | 232,000 |
| | Costless Refuse | 20 % | 200,000 | 0.20[e] | 40,000 | |
| **Pay-or-Tracking Wall** | Tracking | 99 %[c] | 990,000 | 0.24[d] | 237,600 | 270,000 |
| | Pay | 1 % | 10,000 | 3.24[f] | 32,400 | |
| | | | | | **Abs. Difference** | **+ 38,000** |
| | | | | | **Rel. Difference** | **+ 16.4%** |

[a]Assumption: The publisher has one million users, and no user abandons the publisher because of the pay-or-tracking wall.
[b]Assumption: We assume that 80 % of users choose the tracking option in the case of a cookie consent banner.
[c]Result of study 4: A share of approximately 99 % of users choose the tracking option in the case of a pay-or-tracking wall.
[d]Result of study 2: Publishers, on average, generate a revenue of 0.24 EUR per user with the tracking option.
[e]Result and assumption of study 2: Publishers, on average, earn 18% less revenue for users who refuse to be tracked (0.24 EUR x (1-0.18) ≈ 0.20 EUR).
[f]Result of study 2: Publishers charge, on average, 3.24 EUR when bundling the pay option with an ad-free experience.

# Summary of Results and Implications

## Summary of Results

Prestigious news publishers—and more recently, Meta—have started requesting users to pay for privacy by implementing pay-or-tracking walls. While activists and users have raised concerns about privacy becoming a luxury for the affluent, little is known about these pay-or-tracking walls, which prevents a meaningful discussion about their appropriateness.

To address that gap, we conducted a series of empirical studies in order to explore and understand publishers' use of pay-or-tracking walls, users' reactions, and the resulting economic consequences. First, we investigated the popularity and variety of pay-or-tracking walls among publishers, as well as their motivation for pricing the pay option. Secondly, we examined users' reactions by analyzing the impact of introducing a pay-or-tracking wall on publishers' online traffic, along with users' decisions between the pay and tracking options. We then combined the studies' insights to outline the economic consequences of introducing a pay-or-tracking wall.

We found that top European publishers in Austria, France, Germany, and Italy use pay-or-tracking walls. However, the approach was (yet) not apparent in other European countries. The publishers who adopted these walls primarily offered content such as news, lifestyle magazines, cooking, gaming, sports, TV guides, and weather forecasts.

On the topic of design, publishers used a variety of implementations that differently combined the pay option with paid content offerings and advertising-free access. While the tracking option typically included a variety of third-party trackers, the pay option contained substantially less tracking, but was not entirely free of third-party trackers. Notably, we observed that publishers do not base the pay option's prices on the revenue that would otherwise be generated by the tracking option. Instead, publishers seem to follow a different pricing motivation that exceeds the revenue produced via tracking.

Regarding user reactions, publishers did not exhibit a decline in their aggregate online traffic after introducing a pay-or-tracking wall, which indicates that users do not leave and abandon publishers in response. Most users opted for the tracking option and only a few chose the pay option.

Moreover, our results suggest that introducing a pay-or-tracking wall has no adverse economic consequences compared to its direct alternative strategy, a cookie consent banner. For one, the wall seems to have no negative impact on content demand (as represented by publishers' online traffic). For another, the price of the pay option is always higher than the foregone advertising revenue. Thus, publishers can benefit economically from implementing a pay-or-tracking wall because of revenues generated by the pay option and a high share of users who consent to their tracking. In our example calculations, we showed that a pay-or-tracking wall led to 16.4% more revenue for publishers compared to a cookie consent banner.

### Implications and Conclusions

*Implications and Conclusions for Publishers*

Using a pay-or-tracking wall instead of a cookie consent banner is profitable, and setting the pay option's price above the ad revenue when tracking is possible prevents losing money. Moreover, pay-or-tracking walls align with the legal opinions of several national data protection authorities (so long as the price of the pay option is reasonable and not too high). Consequently, publishers should adopt pay-or-tracking walls.

Given the small share of paying users we observed, our studies suggest that users do not have a high willingness to pay to protect their privacy, and the share of paying users is still small. Thus, the prices of the pay options might not be optimal, and publishers will need to experiment with different prices in order to attract more paying users. That said, publishers can choose from a wide range of implementations based on the four dimensions: tracking, advertising, content, and price. Experimenting with other implementations might help to improve the pay option's offering and increase the share of paying users.

*Implications and Conclusions for Users*

While replacing cookie consent banners with pay-or-tracking walls is beneficial for publishers, it leads to a worse situation for users (at least in the short-term), since they can no longer protect their privacy for free. In the long-term, however, users might benefit from the better financing of publishers' content. Thus, pay-or-tracking walls may change the paradigm by asking users to not only "pay with data" or "pay for content," but also "pay for no tracking," stressing that there is "no free lunch." Consequently, online privacy could become a luxury that prompts privacy-sensitive users to focus on a few publishers in order to save money.

In addition to "pay for no tracking," publishers still provide a way to "pay with data" via the tracking option, allowing users to access content or at least parts of it without spending their own money. While "paying for no tracking" may challenge lower-income users, the ability to access content by "paying with data" remains an accessible and barrier-free alternative—albeit at the compromise of privacy.

We want to note the difficulty of comparing the publishers' implementations of pay-or-tracking walls. Moreover, most pay options are not entirely tracking-free, and the payment process also reveals personal data. Thus, users interested in purchasing a pay option should be aware that comparing publishers' offerings requires (substantial) time. Further, users should carefully examine the pay option's tracking policy and be aware that even the pay option discloses user data through its payment process.

*Implications and Conclusions for Regulators*

The implications for regulators are (mostly) the same as for users. However, regulators will want to be attentive to potential user deceptions caused by (i) most pay options not being entirely tracking-free and (ii) publishers' implementations of pay-or-tracking walls being challenging to compare. Consequently, regulators should establish standardized guidelines to specify the design of pay-or-tracking walls, which will help safeguard users' privacy and enhance comparability. Internationally standardized guidelines could also create a level playing field for publishers amid the fragmented privacy landscape shaped by national and regional decisions.

Moreover, regulators should be aware that bundling the pay option with additional features may benefit users and reduce choice complexity, but can lead to publishers charging higher prices for the pay option to compensate for the included benefits. For instance, Meta bundles the "no tracking" feature with "no advertising." However, offering separate features is feasible, so a forced combination of "no tracking" (i.e., pay option) with another feature (e.g., "no advertising") is not required. Thus, Meta could offer a no-tracking alternative at a lower price than it does today.

The low share of users choosing the pay option may stem from prices being too high or users' low willingness to pay for privacy. Therefore, regulators should establish clear pricing guidelines for pay-or-tracking walls, accompanied by transparent criteria to assess the appropriateness and reasonableness of these prices. More broadly, the low share of users opting for the pay option raises the question of whether the European privacy framework improved users' privacy and control of personal data. Pay-or-tracking walls seem to provide the means to expand the practice of user tracking. Policymakers will need to consider whether this aligns with the goals of their legislation.

**Web Appendix: Paying for Privacy: Pay-or-Tracking Walls**

**Web Appendix A: Tracker Categories in the Pay and Tracking Options**

This web appendix provides information about the third-party tracker categories that publishers use in the pay and tracking options. We automatically collected the third-party tracker categories as part of our data collection, using the browser extension Ghostery Insights. Ghostery (2023) provides and maintains the underlying categorization, and describes the purpose and functionality of a tracker, such as advertising, audio/video players, comments, customer interaction, essential, site analytics, or social media.

We first analyzed the distinct number of categories for the pay and tracking options in Figure W1. It shows that the pay and tracking options typically comprise multiple tracker categories. On average, the pay option includes 2.7 different third-party tracker categories, while the tracking option comes with 4.7 third-party tracker categories.

*Figure W1: Distribution of Number of Third-Party Tracker Categories Among Publishers in Pay and Tracking Options*

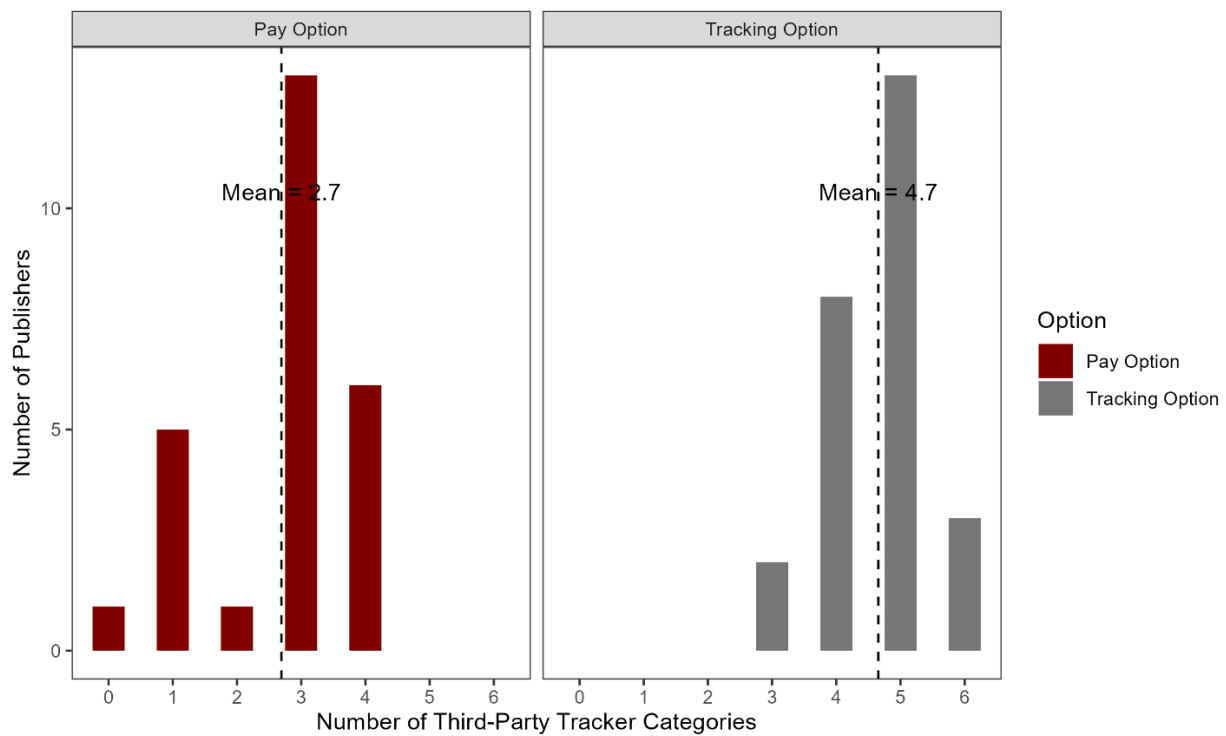

Notes: N = 26 Publishers.

Next, we analyzed the prevalence of specific categories among the publishers' pay and tracking options in Figure W2. It shows that the trackers in the pay option are not used solely for one specific functionality. Instead, the publishers' pay options include third-party trackers related to different functionalities, such as advertising, site analytics, social media, or essential trackers that provide critical website functionality.

*Figure W2: Prevalence of Third-Party Tracker Categories*
*Between Publishers' Pay and Tracking Options*

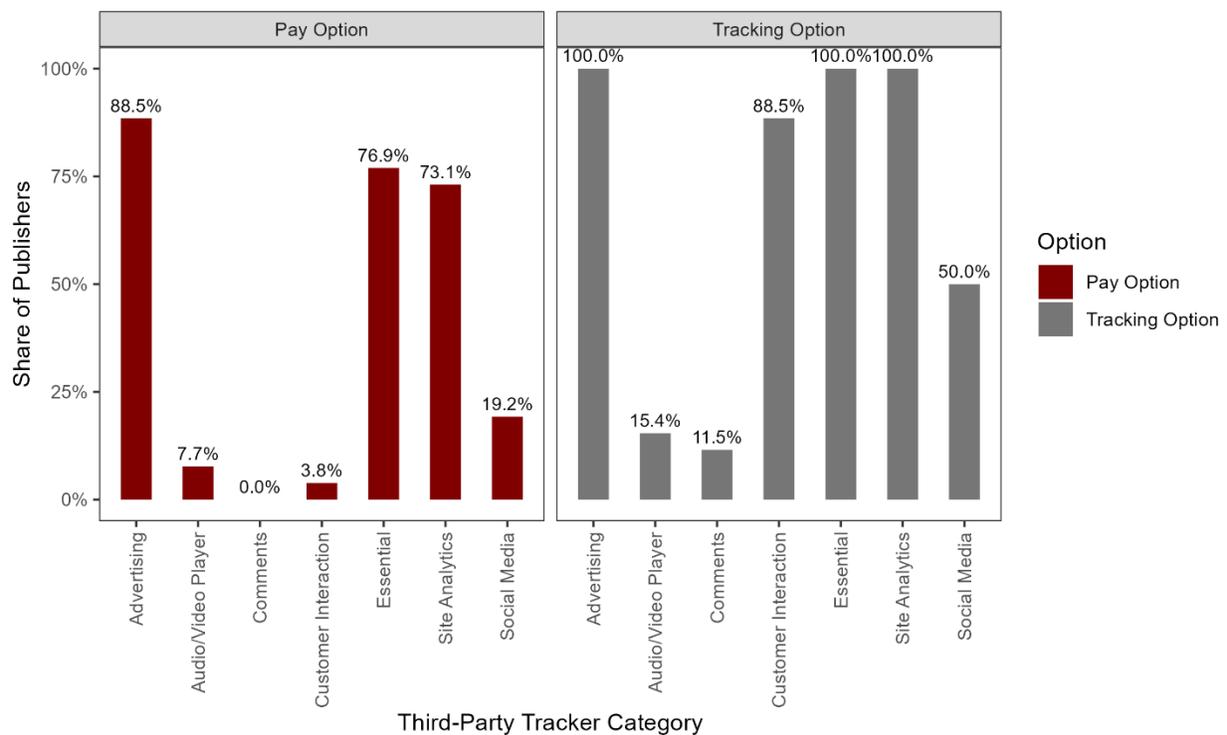

Notes: N = 26 Publishers.

**Web Appendix B: Robustness Test for Alternative Time Horizons and Alternative Dependent Variables Regarding the Impact on Online Traffic**

*Setup of the Robustness Test*

As mentioned in the third empirical study, "Study 3 – Impact on Online Traffic ("Leave Option")", our analysis focused on the impact of introducing a pay-or-tracking wall on online traffic. In this section, we address the concern that our estimation results were affected by the chosen time window (i.e., six weeks before and after the introduction of a pay-or-tracking wall). Therefore, we repeated our analysis from the "Impact of the Introduction of a Pay-or-Tracking Wall on Online Traffic" section using different time windows for the before- and after-periods: specifically, two, four, six, and eight weeks.

Furthermore, we additionally estimated the effect on online traffic using two alternative dependent variables: namely, the number of daily visits and the number of daily page impressions per visit. These two dependent variables are only available for three publishers, as opposed to the seven publishers included in our primary analysis regarding the number of daily page impressions. Footnote 2 in the main manuscript outlines the reason.

*Results of the Robustness Test*

We report the results of our estimation of the coefficient of interest in Table W1. The results confirm the results of our main analysis: Regardless of alternative time windows and dependent variables, we observed no significant decrease in publishers' online traffic.

*Table W1: Results of the Before-After Comparison for Alternative Time Horizons and Alternative Dependent Variables*

| Dependent Variable | Weeks in the Before-Period | Weeks in the After-Period | Estimate of "After" | | |
|---|---|---|---|---|---|
| | | | Model (1) | Model (2) | Model (3) |
| ln(Daily Number of Page Impressions) | 8 | 8 | 0.0150 (0.0285) | 0.0150 (0.0285) | 0.0447 (0.0326) |
| | 8 | 6 | 0.0093 (0.0200) | 0.0093 (0.0200) | 0.0364 (0.0302) |
| | 8 | 4 | 0.0205 (0.0196) | 0.0205 (0.0196) | 0.0392 (0.0292) |
| | 8 | 2 | 0.0341 (0.0258) | 0.0341 (0.0258) | 0.0411 (0.0321) |
| | 6 | 8 | -0.0008 (0.0345) | -0.0008 (0.0345) | 0.0471 (0.0326) |
| | 6 | 6 | -0.0066 (0.0213) | -0.0066 (0.0213) | 0.0376 (0.0296) |
| | 6 | 4 | 0.0046 (0.0130) | 0.0046 (0.0130) | 0.0406 (0.0277) |
| | 6 | 2 | 0.0182 (0.0154) | 0.0182 (0.0154) | 0.0437 (0.0303) |
| | 4 | 8 | -0.0250 (0.0548) | -0.0250 (0.0548) | 0.0484 (0.0351) |
| | 4 | 6 | -0.0308 (0.0395) | -0.0308 (0.0395) | 0.0339 (0.0302) |
| | 4 | 4 | -0.0196 (0.0274) | -0.0196 (0.0274) | 0.0337 (0.0247) |
| | 4 | 2 | -0.0060 (0.0205) | -0.0060 (0.0205) | 0.0375 (0.0261) |
| | 2 | 8 | -0.0457 (0.0799) | -0.0457 (0.0799) | 0.0520 (0.0362) |
| | 2 | 6 | -0.0515 (0.0643) | -0.0515 (0.0643) | 0.0346 (0.0331) |
| | 2 | 4 | -0.0403 (0.0529) | -0.0403 (0.0529) | 0.0288 (0.0288) |
| | 2 | 2 | -0.0267 (0.0441) | -0.0267 (0.0441) | 0.0244 (0.0285) |
| ln(Daily Number of Visits) | 8 | 8 | -0.0598 (0.0242) | -0.0598 (0.0242) | -0.0282 (0.0660) |
| | 8 | 6 | -0.0291 (0.0426) | -0.0291 (0.0426) | -0.0148 (0.0764) |
| | 8 | 4 | 0.0121 (0.0598) | 0.0121 (0.0598) | 0.0073 (0.0791) |
| | 8 | 2 | 0.0473 (0.0747) | 0.0473 (0.0747) | 0.0224 (0.0914) |
| | 6 | 8 | -0.0884 (0.0238) | -0.0884 (0.0238) | -0.0316 (0.0575) |
| | 6 | 6 | -0.0578 (0.0202) | -0.0578 (0.0202) | -0.0191 (0.0661) |
| | 6 | 4 | -0.0165 (0.0337) | -0.0165 (0.0337) | 0.0052 (0.0691) |
| | 6 | 2 | 0.0186 (0.0445) | 0.0186 (0.0445) | 0.0230 (0.0817) |
| | 4 | 8 | -0.1294 (0.0701) | -0.1294 (0.0701) | -0.0334 (0.0453) |
| | 4 | 6 | -0.0988 (0.0531) | -0.0988 (0.0531) | -0.0272 (0.0490) |
| | 4 | 4 | -0.0575 (0.0467) | -0.0575 (0.0467) | -0.0033 (0.0536) |
| | 4 | 2 | -0.0224 (0.0369) | -0.0224 (0.0369) | 0.0192 (0.0686) |
| | 2 | 8 | -0.1848 (0.1453) | -0.1848 (0.1453) | -0.0215 (0.0297) |
| | 2 | 6 | -0.1542 (0.1269) | -0.1542 (0.1269) | -0.0300 (0.0350) |
| | 2 | 4 | -0.1129 (0.1161) | -0.1129 (0.1161) | -0.0194 (0.0474) |
| | 2 | 2 | -0.0778 (0.1007) | -0.0778 (0.1007) | -0.0082 (0.0573) |
| Number of Daily Page Impressions by Daily Visits | 8 | 8 | 0.0652 (0.0329) | 0.0652 (0.0329) | 0.0623 (0.0398) |
| | 8 | 6 | 0.0489 (0.0421) | 0.0489 (0.0421) | 0.0477 (0.0460) |
| | 8 | 4 | 0.0195 (0.0473) | 0.0195 (0.0473) | 0.0200 (0.0487) |
| | 8 | 2 | 0.0110 (0.0499) | 0.0110 (0.0499) | 0.0137 (0.0519) |
| | 6 | 8 | 0.0765 (0.0254) | 0.0765 (0.0254) | 0.0756 (0.0326) |
| | 6 | 6 | 0.0602 (0.0367) | 0.0602 (0.0367) | 0.0606 (0.0401) |
| | 6 | 4 | 0.0308 (0.0433) | 0.0308 (0.0433) | 0.0313 (0.0438) |
| | 6 | 2 | 0.0223 (0.0475) | 0.0223 (0.0475) | 0.0222 (0.0483) |
| | 4 | 8 | 0.0599 (0.0160) | 0.0599 (0.0160) | 0.0574 (0.0311) |
| | 4 | 6 | 0.0437 (0.0277) | 0.0437 (0.0277) | 0.0448 (0.0360) |
| | 4 | 4 | 0.0143 (0.0353) | 0.0143 (0.0353) | 0.0175 (0.0377) |
| | 4 | 2 | 0.0058 (0.0410) | 0.0058 (0.0410) | 0.0044 (0.0435) |
| | 2 | 8 | 0.0331 (0.0370) | 0.0331 (0.0370) | -0.0001 (0.0047) |
| | 2 | 6 | 0.0169 (0.0413) | 0.0169 (0.0413) | -0.0060 (0.0225) |
| | 2 | 4 | -0.0125 (0.0497) | -0.0125 (0.0497) | -0.0249 (0.0326) |
| | 2 | 2 | -0.0211 (0.0603) | -0.0211 (0.0603) | -0.0385 (0.0428) |

Significance levels: * $p < 0.05$, ** $p < 0.01$, *** $p < 0.001$.

Standard errors are clustered at the publisher level and reported in brackets.

Notes: This table shows the coefficient (After) from the OLS regressions of the before-after comparison for different time windows and dependent variables. Model (1) refers to model (1) in Table 4, which considers only publisher fixed effects. Instead of publisher fixed effects, model (2) considers publisher-specific day-of-week fixed effects. Model (3) includes publisher-specific day-of-week fixed effects and users' news interest. None of the estimated "after"-coefficients significantly differed from zero when applying the indicated significance levels.

# Web Appendix C: Robustness Test for Alternative Identification Strategies Regarding the Impact on Online Traffic

*Setup of the Robustness Test*

As mentioned in the third empirical study, "Study 3 – Impact on Online Traffic ("Leave Option")", our analysis focused on the impact of introducing a pay-or-tracking wall on online traffic. This section addresses the concern that the before-after comparison may not completely capture time trends. Therefore, we performed a robustness check by using an alternative identification strategy.

We used the synthetic difference-in-differences estimator, introduced by Arkhangelsky et al. (2021), to capture time trends based on a synthetic control group of comparable publishers. We compared each publisher who introduced a pay-or-tracking wall with its synthetic representation if the publisher had not introduced a pay-or-tracking wall. The synthetic representation builds upon other comparable control publishers; its composition is estimated such that it follows a similar time trend in the before-period. We then used this synthetic control to calculate a weighted difference-in-difference estimator.

*Definition of the Dependent Variable for Online Traffic*

We aggregated the number of daily page impressions for each publisher to a weekly level to reduce the variability induced by day-of-the-week effects. Further, we applied a log transformation, so that the log of the number of weekly page impressions served as our dependent variable.

*Sample of Publishers with Pay-or-Tracking Wall*

We analyzed the same sample of publishers as in the before-after analysis (i.e., the seven Top-50 German publishers who introduced a pay-or-tracking wall). Those publishers primarily offer national news to their users.

*Construction of Synthetic Control*

For each individual publisher, we estimated the impact of introducing a pay-or-tracking wall on online traffic. Thus, we also constructed an individual synthetic control for each publisher: specifically, a weighted combination of a donor pool of comparable publishers. The weighted combination was designed to match the publisher before introducing its pay-or-tracking wall. We chose a before-period of 24 weeks to ensure a sufficiently long time window and form a reliable estimation of the synthetic representation. We used an after-period of six weeks, in line with our primary analysis. Below, we elaborate on the donor pool of comparable publishers.

*Donor Pool for the Synthetic Control Groups*

This method necessitates that we choose a donor pool consisting of comparable publishers. Thus, we built synthetic control groups based on publishers that are similarly exposed to concurrent events and time trends. As our sample consists of publishers that provide national news, the donor pool should include publishers whose online traffic also follows national news events. Publishers that do not provide national news are likely not following similar time trends.

Many Top-50 German publishers that provide national news have already introduced a pay-or-tracking wall. For instance, the German Audit Bureau of Circulation (our data source) only included two Top-50 publishers who provide national news but did not yet introduce a pay-or-tracking wall. We placed these two publishers in our donor pool.

Further, we utilized the different adoption times of the publishers' pay-or-tracking walls. We assumed that the potential impact of introducing a pay-or-tracking materializes after 180 days, after which the publisher's online traffic stabilizes at its potential new level ("adjustment period"). We included other publishers who introduced a pay-or-tracking wall in the donor pool if the "adjustment period" did not overlap with the focal publisher's estimation period, involving the 24 weeks before introducing a pay-or-tracking wall and six weeks after.

Thus, each publisher's donor pool consists of (a) two national news publishers who did not introduce a pay-or-tracking wall and (b) the other publishers who introduced a pay-or-tracking wall but did not experience an overlap between their "adjustment period" and the focal publisher's estimation period. The latter group encompassed the seven national news publishers who introduced a pay-or-tracking wall. Further, we included another publisher who introduced a pay-or-tracking wall and covers national news, but for which we had missing data around the introduction period but not for other periods.

In total, the donor pool comprises ten publishers. If none of the "adjustment periods" overlap with the "estimation period", then our estimation would include the focal publisher for whom we estimated the impact of introducing a pay-or-tracking and nine control publishers out of the donor pool.

### Results of the Robustness Test

The synthetic difference-in-differences method incorporated an estimation equation similar to the one used for the difference-in-difference approach, with the major difference being the use of a synthetic control as the "non-treated" comparison:

$$\ln(\# \ Weekly \ Page \ Impressions_{iw})$$
$$= \alpha + \beta After_w + \gamma Treatment_i + \delta(Treatment_i * After_w) + \epsilon_{iw} \quad (1)$$

Equation (1) includes an indicator variable $After_w$, which takes the value 1 for all weeks $w$ after introducing the pay-or-tracking wall. The indicator variable $Treatment_i$ is 1 for the focal publisher $i$ who introduced a pay-or-tracking wall, and 0 for its synthetic control. The coefficient of interest is $\delta$, which covers the interaction between the two indicator variables $After_w$ and $Treatment_i$.

We present the summary of each publisher's estimation in Table W2, which shows that the coefficient of interest is never statistically different from zero. This result suggests that none of the publishers experienced a decline in their online traffic, thereby confirming the results of our primary analysis via the before-after comparison.

*Table W2: Results of Publisher-Specific Synthetic Difference-in-Difference Estimations*

| Publisher | Number of Publishers in Donor Pool | Number of Observations | After x Treatment |
|---|---|---|---|
| Publisher 1 | 7 | 240 | -0.0474 (0.0569) |
| Publisher 2 | 8 | 270 | 0.0663 (0.0613) |
| Publisher 3 | 6 | 210 | -0.0229 (0.0267) |
| Publisher 4 | 8 | 270 | 0.0151 (0.0690) |
| Publisher 5 | 6 | 210 | 0.0162 (0.0206) |
| Publisher 6 | 6 | 210 | 0.0609 (0.0563) |
| Publisher 7 | 8 | 270 | -0.0545 (0.0757) |